\documentclass[a4paper,11pt]{article}
\pdfoutput=1
\usepackage{jheppub}
\usepackage{subfigure}

\usepackage{slashed}
\usepackage{bm,wasysym}
\usepackage[normalem]{ulem}

\usepackage{amsmath}
\usepackage{amssymb,dsfont,bbold}

\usepackage{feynmp}

\usepackage{placeins}

\newcommand{\dd}{{\rm d}}
\def\slash#1{\setbox0=\hbox{$#1$}\dimen0=\wd0
      \setbox1=\hbox{/} \dimen1=\wd1 \ifdim\dimen0>\dimen1
      \rlap{\hbox to \dimen0{\hfil/\hfil}} #1                        \else
      \rlap{\hbox to \dimen1{\hfil$#1$\hfil}}
      /   \fi}

\newcommand{\lsim}{
\mathrel{\hbox{\rlap{\hbox{\lower4pt\hbox{$\sim$}}}\hbox{$<$}}}}

\newcommand{\gsim}{
\mathrel{\hbox{\rlap{\hbox{\lower4pt\hbox{$\sim$}}}\hbox{$>$}}}}

\allowdisplaybreaks[2]

\newcommand{\para}{\parallel}

\title{QCD Factorization for $\boldmath B \to \pi\pi\ell\nu$ Decays \newline at Large Dipion Masses}

\author[1]{Philipp B\"oer,}
\author[2]{Thorsten Feldmann,}
\author[1,2]{Danny van Dyk}

\affiliation[1]{Theoretische Physik 1, 
Universit\"at Siegen, Walter-Flex-Stra\ss{}e 3, D-57068 Siegen, Germany}
\affiliation[2]{Physik-Institut, Universit\"at Z\"urich, Winterthurerstrasse 190, CH-8057 Z\"urich, Switzerland}

\emailAdd{boeer@physik.uni-siegen.de}
\emailAdd{thorsten.feldmann@uni-siegen.de}
\emailAdd{dvandyk@physik.uzh.ch}

\abstract{We introduce a factorization formula for 
semi-leptonic $b \to u$ transitions in the exclusive
decay mode $B^- \to \pi^+\pi^-\ell^-\bar\nu_\ell$ in 
the limit of large pion energies and large dipion invariant mass.
One contribution can be described
in terms of a universal $B \to \pi$ form factor and the convolution of a
short-distance kernel $T^{\rm I}$ with the respective
light-cone distribution amplitudes
(LCDAs) of the positively charged pion. 
The second contribution, at leading power,
completely factorizes, with
a short-distance kernel $T^{\rm II}$  convoluted with the leading-twist LCDAs for 
both pions and the $B$-meson.
We calculate the leading contributions to the short-distance
kernels $T^{\rm I}$ and $T^{\rm II}$ in fixed-order perturbation theory,
and discuss the approximate relations among the resulting
$B \to \pi\pi$ partial-wave form factors.
Our results provide useful theoretical constraints 
for phenomenological models that aim to analyze the complete $B \to \pi\pi \ell\nu$
phase space.
}


\note{\today.
Siegen:  SI-HEP-2016-05,  QFET-2016-02; Zurich: ZU-TH 19/16; EOS-2016-04}

\begin{document}

\maketitle


\clearpage 

\section{Motivation}

Exclusive charmless $B$-meson decays play an important role for the 
phenomenological analysis of quark flavour transitions 
in the Standard Model (SM) or its possible new-physics (NP) extensions
(see e.g.\ the reviews in \cite{Buchalla:2008jp,Antonelli:2009ws,Bediaga:2012py,Buras:2013ooa,Bevan:2014iga}).
On the theoretical side, a systematic separation of 
short-distance effects in Quantum Chromodynamics (QCD) and 
long-distance hadronic physics can (at least partially) be 
achieved by utilizing an expansion in inverse powers of the large 
$b$-quark mass, i.e.\ $\Lambda/m_b\ll 1$, where $\Lambda$ is a typical 
hadronic scale, $\Lambda \lesssim 1$~GeV.
In particular, this can be used to derive factorization formulas
that allow one to implement QCD radiative corrections 
to the ``naive'' factorization approximation on a field-theoretical
basis.

Factorization theorems for charmless nonleptonic $B$-meson decays 
into two mesons have been established at leading power in the heavy-mass
expansion \cite{Beneke:1999br,Beneke:2001ev}. 
Higher-order perturbative corrections have been calculated in
\cite{Bell:2007tv,Bell:2009nk,Beneke:2009ek,Bell:2015koa}
and \cite{Beneke:2005vv,Kivel:2006xc,Pilipp:2007mg,Beneke:2006mk}
(see also \cite{Feldmann:2014iha} for a brief overview).
One of the main motivations in this context was to 
increase the precision of theoretical predictions or, at least, 
get a more reliable theoretical assessment of hadronic uncertainties,
which cannot be described by simple quantities like decay constants or 
transition form factors; see e.g.\ the phenomenological 
analyses in \cite{Beneke:2003zv,Beneke:2006hg}.
The energies of the light hadrons in exclusive $B$-meson
decays are not extremely large and power corrections still
provide a major source of hadronic uncertainties, which are difficult to
estimate and thus obscure the NP sensitivity in exclusive $B$-meson decays
(see e.g.\ \cite{Li:2005kt,Cheng:2004ru,Bauer:2005kd,Feldmann:2008fb}). 
For transitions which are dominated
by tree-level exchange of $W$-bosons in the SM,
potential NP effects are expected to play a subdominant role.
The non-leptonic case has been extensively studied in the past;
see e.g.\ the recent discussion in \cite{Bell:2009fm} and references therein.
In this work, we focus on the semileptonic decays 
$B^- \to \pi^+\pi^-\ell^-\bar\nu_\ell$,
which are induced by
$b \to u \ell^- \bar \nu_\ell$ transitions and, in the SM, only involve one
effective operator containing the left-handed $b \to u$ quark current. QCDF is expected to be
applicable in the kinematic situation where, in the $B$-meson rest frame,
both pions recoil against each other with large energies of order $m_b/2$. 
The theoretical description features elements known from the analysis of 
nonleptonic $B \to \pi\pi$ decays as in  \cite{Beneke:2003zv,Beneke:2006hg}
and semileptonic $B \to \pi\ell\nu$ decays \cite{Beneke:2000wa} and 
leads to a very similar QCD factorization formula. The confirmation of this
factorization formula by explicit calculation of the leading non-trivial
contributions to the hard-scattering kernels is the main subject of this
paper. However, we will not aim at a rigorous factorization proof 
within the context of Soft-Collinear Effective Theory \cite{Bauer:2000yr,Beneke:2002ph};
a discussion along the lines of \cite{Beneke:2003pa,Beneke:2003xr} is left for future work.

One advantage of the $B \to \pi\pi\ell\nu$ decay compared to its 
non-leptonic counterpart $B \to \pi\pi$ is its richer kinematic structure that
opens the possibility to
analyze the angular distribution in the 4-body final state. Similar angular analyses
have also been successfully exploited in phenomenological studies
for other multi-body decay modes like $B \to K\pi\ell\ell$ \cite{Kruger:2005ep,Altmannshofer:2008dz,Bobeth:2010wg},
$B_s \to K \pi \ell \nu$ \cite{Feldmann:2015xsa,Meissner:2013pba}, and 
$\Lambda_b \to N\pi \ell \ell$ \cite{Boer:2014kda}.
In particular, in certain corners of the phase space one finds 
approximate form factor relations that lead to simple 
theoretical predictions in the limit $m_b \to \infty$ \cite{Faller:2013dwa}.
As we will see, this will also be the case in the kinematic situation 
that we are considering in this work.
It could thus be interesting to
interpolate between different phase-space regions in $B \to \pi\pi\ell\nu$
decays, using the results of this work and others 
(see e.g.\ \cite{Kang:2013jaa,Meissner:2013hya,Hambrock:2015aor}).
Our formalism can also be generalized to certain 
phase-space regions in multi-pion final states, as considered
in \cite{Krankl:2015fha}.

Our paper is organized as follows.
In the next section we start with a brief summary of the 
relevant kinematic variables and the power-counting scheme
that underlies the QCD factorization formula for $B^-\to\pi^+\pi^-\ell^- \bar\nu_\ell$
that will be 
investigated in Section~\ref{sec:fac}. In that section,
we give a detailed derivation of the leading contribution
(i.e.\ ${\cal O}(\alpha_s)$) to the kernel $T^{\rm I}$,
also including contributions from the 
twist-3 distribution amplitudes of the positively charged 
pion, which are formally of subleading power but numerically
enhanced.  
Furthermore, we calculate the kernel $T^{\rm II}$, which arises
from spectator scattering.
We identify the endpoint-divergent contributions, which 
will be shown to exactly match the corresponding terms that appear 
in the universal ``soft'' $B \to \pi$ form factor. The remaining
finite terms provide the ``factorizable'' corrections of 
order $\alpha_s^2$ to the $B \to \pi\pi$ form factors at 
large dipion mass. In Section~\ref{sec:obs} we discuss the 
phenomenological implications, on the one hand in terms
of approximate relations between the individual
$B \to \pi\pi$ partial-wave form factors, 
and on the other hand in terms of numerical
estimates for two observables: the integrated decay rate
and the pionic forward-backward asymmetry, in bins of the invariant
dilepton and dipion masses.
We conclude with a brief summary in Section~\ref{sec:summ}.
Detailed information on our conventions for the definition
of the dipion form factors, as well as on the calculation of
the individual diagrams contributing to the kernel $T^{\rm II}$
are collected in two appendices.

\section{Kinematics and Power Counting}

\label{sec:pc}

We define the kinematics for the decay 
$$
 B^-(p) \to \pi^+(k_1) \, \pi^-(k_2) \, \bar \nu_\ell(q_1) \, \ell^-(q_2)
$$
following the conventions in \cite{Faller:2013dwa}.
In the kinematic regime that we are interested in, it is safe to 
neglect the pion mass compared to the large $B$-meson mass and 
pion energies at large hadronic recoil. We will therefore set $M_\pi^2 \to 0$
throughout the paper.
Defining the sums and differences of hadronic and leptonic
momenta as
\begin{align}
q = q_1+q_2 \,, & \qquad k= k_1+k_2 \,,
\cr 
\bar q= q_1-q_2 \,, & \qquad \bar k= k_1-k_2 \,,
\end{align}
the hadronic system can be described by three kinematic
Lorentz invariants, which can be chosen as the momentum
transfer $q^2$, the dipion invariant mass $k^2$, and the 
scalar product
\begin{align}
 q \cdot \bar k =  \frac{\sqrt{\lambda}}{2} \, \cos\theta_\pi \,.
\end{align}
Here $\theta_\pi$ refers to the polar angle of the $\pi^+$ meson in the 
dipion rest frame, and 
\begin{align}
 \lambda \equiv 
 M_B^4+q^4+k^4-2 \, (M_B^2 q^2 + M_B^2 k^2 + q^2 k^2)
\end{align}
 is the K\"all\'en function.
For the following discussion it is sometimes 
 more convenient to use the independent variables
 \begin{align}
E_{1,2} \equiv \frac{p \cdot k_{1,2}}{M_B} = \frac{M_B^2 + k^2 - q^2 \pm \cos\theta_\pi \sqrt{\lambda}}{4 M_B}
\qquad \mbox{and} \qquad k^2 \,, 
\label{kin2}
\end{align}
 where $E_{1,2}$ denote the energies of
 the individual pions in the $B$-meson rest frame, with 
\begin{align}
q^2 &= M_B^2 - 2 M_B \, (E_1+E_2) + k^2 \,, 
\qquad 
q \cdot \bar k  = M_B \, (E_1-E_2) \,,
\label{kinrel}
\end{align}
and
\begin{align}
\lambda & = 4 M_B^2 \left( (E_1+E_2)^2-k^2 \right) \,.
 \label{kin3}
 \end{align}
The power counting that underlies the
 factorization formula, to be introduced below, 
follows from the requirements that:
\begin{itemize}
\item[(i)] The energies of both pions in the $B$-meson rest frame 
are large to allow for the factorization
of soft modes in the $B$-meson and collinear modes in the pions,
\begin{align}
 E_{1,2} \gg \Lambda \,,
 \end{align}
where $\Lambda$ is a typical hadronic scale;
\item[(ii)] The invariant mass of the dipion system $k^2$ is large, 
in order to allow for the factorization of collinear modes
in the two different pion directions:
\begin{align}
    k^2 \gg \Lambda^2\,.
\end{align}
 \end{itemize}
Allowing for generic values of $q^2$, $k^2$ and $|\cos\theta_\pi|$, the minimal
pion energy corresponds to
\begin{align}
    \label{eq:Emin}
    E_{1,2} \geq E_{\rm min}(q^2, k^2, |\cos\theta_\pi|)
        = \frac{M_B^2 + k^2 - q^2 - |\cos\theta_\pi| \sqrt{\lambda}}{4 M_B}\,.
\end{align}
Criterium $(i)$ is therefore fulfilled if $E^{\rm min} \gg \Lambda$.
%
For a quantitative estimate, we also have to take into account that
the ratio $M_B/\Lambda$ is not extremely large, and thus choose the
phase space boundaries carefully. 
A conservative benchmark case would be, for instance, to require 
$E_{\rm min} = M_B/3 \simeq 1.76~{\rm GeV}$.
Without any additional cuts on $|\cos\theta_\pi|$ and regardless of the
value of $q^2$, this can be achieved 
by setting $k^2_{\rm min} = 2 M_B^2/3$ (see App.~\ref{app:kinematics}). This defines
\begin{align}
  \mbox{Scenario A:} \quad &
  k^2_{\rm min} = 2M_B^2/3 \simeq 18.6~{\rm GeV}^2 
  \cr \quad \Rightarrow \quad &E_{\rm min} =  M_B/3 \simeq 1.76~{\rm GeV} \qquad (\mbox{for $|\cos\theta_\pi|\leq 1$}).
\end{align}
Notice that in this case one finds that $|E_1-E_2| \leq 0.9$~GeV, i.e.\ one
is very close to the kinematic endpoint, where
\begin{align*}
   k^2 \simeq (E_1+E_2)^2 \sim M_B^2 \,,  \qquad |E_1-E_2| \sim \Lambda \ll M_B
   \,, \qquad \sqrt{\lambda} \ll M_B^2 \,.
\end{align*}
For $q^2 \to 0$ this includes the special case for the kinematics in
non-leptonic $B \to \pi \pi$ decays \cite{Beneke:1999br}.
In a still reasonable
benchmark scenario we allow for slightly smaller values of $E_{\rm min}$, 
which can be achieved (again for all values of $q^2$ and $|\cos\theta_\pi|$)
by a somewhat relaxed bound on $k^2$,
ending up with
\begin{align}
  \mbox{Scenario B:} \quad &
  k^2_{\rm min} = M_B^2/2 \simeq 13.9~{\rm GeV}^2 \,,
  \cr \quad \Rightarrow\quad & E_{\rm min} = M_B/4 \simeq 1.32~{\rm GeV} & \qquad  (\mbox{for $|\cos\theta_\pi|\leq 1$}).
\end{align}
The range of $k^2$ can be further extended by restricting the size of $|\cos\theta_\pi|$, which
yields a non-trivial lower-bound on the size of $k^2$. For the case considered in the following,
the bound reads
\begin{align}
    E_{\rm min} < \frac{\sqrt{a^2 - 1}}{2a}\,\sqrt{k_{\rm min}^2}\,,
\end{align}
where $|\cos\theta| \leq 1/a$. (Further details and the derivation of this bound are
relegated to App.~\ref{app:kinematics}.)
Aiming, as an example, at a value $k^2_{\rm min} = M_B^2/4$ for 
an angular bound $|\cos\theta_\pi|\leq 1/3$, we obtain
\begin{align}
    \mbox{Scenario C:} \quad & k^2_{\rm min} = M_B^2/4 \simeq 7~{\rm GeV}^2 \,, \quad |\cos\theta_\pi|\leq 1/3
  \cr \quad \Rightarrow \quad & 
    E_{\rm min} = \frac{1}{3\sqrt{2}} M_B \simeq 1.24~{\rm GeV}\,.
\end{align}
This includes the so-called 
``mercedes-star'' configuration in $B \to 3\pi$ decays \cite{Krankl:2015fha},
for which $E_1=E_2=M_B/3$, $k^2 = M_B^2/3$ and $\cos\theta_\pi=0$.

Note that in each scenario above,
the maximal value of the momentum transfer is given by 
$$q^2_{\rm max} = (M_B -\sqrt{k^2_{\rm min}})^2\,,$$
such that
$$\frac{q^2_{\rm max}}{M_B^2} \simeq 0.03\quad\text{(Scenario A)}\,,
\qquad\frac{q^2_{\rm max}}{M_B^2} \simeq 0.09\quad\text{(B)}\,,
\qquad\frac{q^2_{\rm max}}{M_B^2} \simeq 0.25\quad\text{(C)}.$$
In the following, we will retain the entire $q^2$-dependence in the theoretical expressions.
In Scenarios A and B, however, the numerical values of $q^2$ are sufficiently small 
that one can approximate the results by only keeping the linear term of
a Taylor expansion in $\sqrt{q^2}/M_B$.

\section{Factorization Formula}

\label{sec:fac}

In the limit where the two final-state pions in the $B$-meson rest frame
move nearly back-to-back with large energy and large invariant mass,
the hadronic matrix elements for generic $b\to u$ currents in the SM or beyond are expected
to factorize in a similar way as the hadronic matrix elements of 4-quark and chromomagnetic
penguin operators appearing in non-leptonic $B\to \pi\pi$ decays \cite{Beneke:1999br,Beneke:2001ev}.
The noticeable difference between the two cases stems from the fact that the
perturbative expansion for the short-distance kernels in $B \to \pi\pi\ell\nu$
requires at least one hard gluon exchange to generate the additional quark-antiquark
pair ending up in the final-state pions.
We thus introduce the following factorization formula
\begin{align}
 &\langle \pi^+(k_1)\pi^-(k_2)|\bar \psi_u \, \Gamma \, \psi_b |B^-(p)\rangle 
\cr 
&= \frac{2\pi \, f_\pi}{k^2} \left\{ \xi_{\pi}(E_2;\mu) \
 \int_0^1 du \, \phi_\pi(u;\mu) \, T_\Gamma^{\rm I}(u,k^2,E_1,E_2;\mu) \right.
 \cr 
 & \left. \phantom{\frac{2\pi \, f_\pi}{k^2} } \qquad + 
 \frac{\pi^2 f_B f_\pi M_B}{N_C E_2^2} \,
\int_0^1 du \int_0^1 dv \int_0^\infty \frac{d\omega}{\omega} \right.
\cr &  \left. \phantom{\frac{2\pi \, f_\pi}{k^2} }\qquad \quad \, \times 
 \, \phi_\pi(u;\mu) \, \phi_\pi(v;\mu) \, \phi_B^+(\omega;\mu) \, 
  T_\Gamma^{\rm II}(u,v,\omega,k^2,E_1,E_2;\mu) \right\}
 & \cr 
 & \qquad + \mbox{power corrections} \,.
 \label{theorem}
\end{align}
In the first term, $\xi_\pi(E_2)$ denotes the universal
non-factorizable (``soft'') $B^- \to \pi^-$ form factor in SCET \cite{Beneke:2000wa,Beneke:2002ph,Beneke:2003pa}, which can be defined as
\begin{align}
 \langle \pi^-(k_2) | \bar \xi^{(u)} \, \Gamma_X \, h_v^{(b)} |B(v)\rangle 
 &= \xi_\pi(E_2) \, {\rm tr} \left[\slashed k_2 \, \Gamma_X \, P_v \right] \,. 
 \label{xipi}
\end{align} 
Here 
\begin{align}
P_v \equiv \frac{\slashed p + M_B}{2 M_B} \simeq \frac{1+\slashed v_b}{2}
\end{align}
is the usual projector on the large components $h_v^{(b)}$ of the heavy-quark spinor
in Heavy-Quark Effective Theory (HQET) with the heavy-quark velocity $v_b^\mu$.  
Furthermore,
$\xi^{(u)}$ denotes the large component of an energetic
up-quark spinor field in SCET. 
Finally,
$\phi_\pi(u)$ is the leading-twist LCDA of the (in this case positively charged) pion, 
and $T_\Gamma^{\rm I}$
denotes the short-distance kernel from hard gluon interactions with the constituents
of the pions in the final state. The second term factorizes completely into 
leading-twist LCDAs, $\phi_\pi$ and $\phi_B$, for the pions and the $B$-meson, 
convoluted with a short-distance kernel that
contains the contributions from  hard-collinear gluon exchange
with the (would-be) spectator quark in the $B$-meson as well as 
additional hard-gluon corrections. (The normalization factors in (\ref{theorem})
have been chosen for convenience.)

In the following, we are going to confirm this factorization 
structure by explicit calculation of the leading contributions
to the kernels $T^{\rm I}$ and $T^{\rm II}$.

\begin{figure}[t]
\begin{center}
\includegraphics[width=0.98\textwidth]{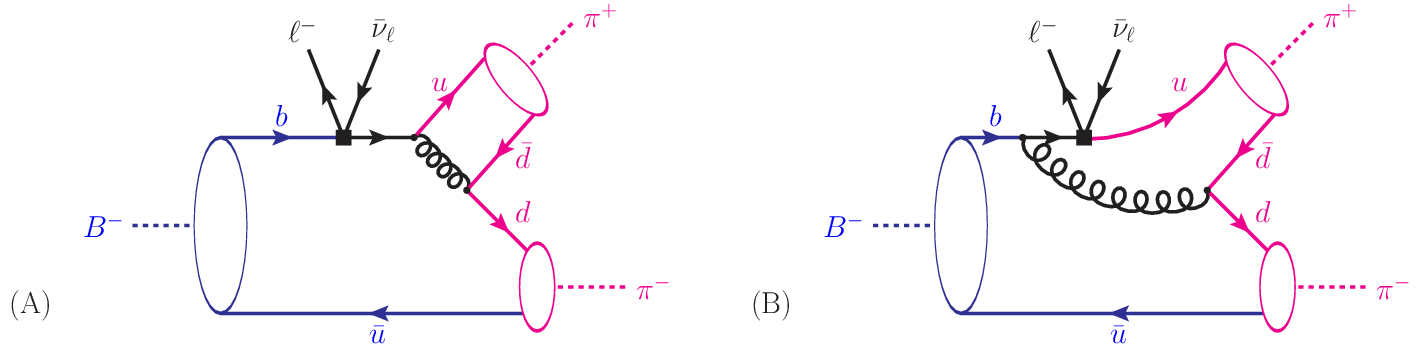}
\end{center}
\caption{\label{fig1} Sketch of QCD factorization in
$B^- \to \pi^+\pi^-\ell^-\bar \nu_\ell$ decays at large dipion mass: Diagrams (i) and (ii)
show the leading decay mechanism from hard gluon exchange.
Radiative corrections, including factorizable and non-factorizable
spectator interactions (see below) are not shown. (The colour coding refers to 
soft momentum modes in blue, and collinear momentum modes in magenta.)}
\end{figure}

\subsection{The kernel $T^{\rm I}$}
\label{sec:T1}
The kernel $T^{\rm I}$ contains the short-distance QCD effects 
that \emph{do not} involve the spectator quarks (and gluons)
in the $B$-meson. The non-trivial tasks are then to show that
\begin{enumerate}
  \item the leading-power contributions indeed only involve  
    the leading-twist pion distribution amplitude of the $\pi^+$ meson,
  \item additional spectator interactions that would formally 
   lead to endpoint-divergences in $T^{\rm II}$ are indeed
   universal and can be absorbed into the soft form factor $\xi_\pi$.
\end{enumerate}
We are going to address the first issue in this subsection by computing the 
leading amplitude term for the semi-partonic process $b \to \pi^+ d \, \ell^-\bar\nu_\ell$.
The second problem is left for the next subsection when we discuss
the leading spectator-scattering diagrams. We stress that, at this point, 
we are neither aiming at an all-order proof of the factorization formula, 
nor at its formal embedding into SCET. 

At leading order in the strong-coupling constant, and projecting
onto the 2-particle Fock state for the energetic pion, the process
$b \to \pi^+ d \, \ell^- \bar\nu_\ell$ is described by the two diagrams in Fig.~\ref{fig1}.
The leading-twist momentum space projector for the final-state pion 
(see e.g.\ \cite{Beneke:2000wa}), reads
\begin{align}
\label{eq:Mpi+}
 {\cal M}_{\pi^+}^{(2)}(u) &= i f_\pi \, \frac{\mathds{1}}{N_C} \, \frac{\slashed k_1 \gamma_5}{4} \, \phi_\pi(u)
 \,, \qquad \mbox{[$(k_1)^2=0$]}
\end{align}
where $u$ and $\bar u =1-u$ are the longitudinal momentum fractions
of the quark and anti-quark in a 2-particle Fock state, i.e.\
\begin{align}
 k_{q1}^\mu & \simeq u \, k_1^\mu \,, \qquad 
 k_{\bar q1}^\mu  \simeq \bar u \, k_1^\mu \,.
\end{align}
Using Eq.~(\ref{eq:Mpi+}),
one obtains for a generic Dirac matrix $\Gamma$
\begin{align}
\langle \pi^+(k_1) \, d(k_{q2}) | \bar \psi_u \Gamma \psi_b|b(p_b)\rangle 
&= 4\pi \alpha_s \, C_F \, \int_0^1 du  \left[ \bar u(k_{q2}) \, \Gamma_X \, u(p_b) \right]
\end{align}
with 
\begin{align}
\Gamma_X & =  - \frac{\gamma_\alpha  \, {\cal M}_{\pi^+}^{(2)}(u) \, \gamma^\alpha  \,
 (\slashed p_b -\slashed q) \, \Gamma}{(p_b-q)^2 \, (p_b - q -  u k_1)^2} 
- \frac{\gamma_\alpha  \, {\cal M}_{\pi^+}^{(2)}(u) \, \Gamma \, (u \slashed k_1+\slashed q + m_b)
 \, \gamma^\alpha  }{[(u k_1 + q)^2 -m_b^2]\, (p_b - q - u k_1)^2} \,,
\label{bpid}
\end{align}
in Feynman gauge.\footnote{One should not confuse the momentum fractions $u,\bar u=1-u$ with
the on-shell Dirac spinors $u(p),\bar u(p)$.}
Here we have used momentum conservation to replace
$k_{q2}^\mu= p_b^\mu - q^\mu - k_1^\mu$.
In the heavy-quark limit, we can further approximate $m_b \simeq M_B$, 
and $p_b^\mu \simeq p^\mu$, such that 
the denominators of the propagators can be expressed in terms of
the hadronic Lorentz invariants defined above,
\begin{align}
(p_b-q)^2 & \simeq (p-q)^2 = k^2 \,, \cr 
(p_b-q - u k_1)^2 &\simeq (k_2 + \bar u k_1)^2 = \bar u  k^2 \,,
\cr 
(u  k_1 + q)^2 -m_b^2 & \simeq  (p - \bar u k_1 - k_2)^2 - M_B^2 
= \bar u \left(k^2- 2  M_B E_1 \right) - 2 M_B E_2 \,.
\end{align}
Assuming the Feynman mechanism to work, i.e.\ all endpoint-divergences
from hard-collinear spectator scattering can be absorbed into the universal
form factor $\xi_\pi$ (which will be shown
by explicit calculation of $T_\Gamma^{\rm II}$ below), we can 
replace the semi-partonic amplitude (\ref{bpid}) by the hadronic one via 
(\ref{xipi}),
\begin{align}
\langle \pi^+(k_1)\pi^-(k_2)|\bar \psi_u \, \Gamma \, \psi_b |B^-(p)\rangle
&= 
4\pi \alpha_s  C_F \, \xi_\pi(E_2) \, \int_0^1 du \,
{\rm tr} \left[ \slashed k_2 \, \Gamma_X \, P_v \right] \,.
\label{me1}
\end{align}
From this we can read off the LO contribution to the 
hard-scattering kernel for a given Dirac structure $\Gamma$.
For the presentation of the results, we find it
convenient to define a basis of Dirac traces,\footnote{The corresponding structures without $\gamma_5$ do not appear
due to parity invariance of QCD.}
\begin{align}
  s_1 &\equiv {\rm tr}[ \slashed k_1 \gamma_5 \Gamma P_v] \,, &&
   s_2 \equiv {\rm tr}[ \slashed k_2 \gamma_5 \Gamma P_v] \,,
   \cr 
   s_3 &\equiv  {\rm tr}[ \slashed k_1 \gamma_5 \Gamma ] \,, &&
   s_4 \equiv  {\rm tr}[ \slashed k_2 \gamma_5 \Gamma ] \,,
   \cr 
   s_5 &\equiv \frac{1}{M_B} \, {\rm tr}[ \slashed k_2 \slashed k_1 \gamma_5 \Gamma P_v ] \,, &&
   s_6 \equiv  \frac{1}{M_B} \, {\rm tr}[ \slashed k_1 \slashed k_2 \gamma_5 \Gamma P_v ] \,,
   \cr 
   s_7 &\equiv \frac{1}{M_B} \, {\rm tr}[ \slashed k_2 \slashed k_1 \gamma_5 \Gamma ] \,, &&
   s_8 \equiv  \frac{1}{M_B} \, {\rm tr}[ \slashed k_1 \slashed k_2 \gamma_5 \Gamma ] \,.
   \label{sdef}
\end{align}
(Notice that in case of vector and axial-vector currents, one has $s_3=2 s_1$, 
$s_4 = 2 s_2$, and $s_7=s_8=0$.)
In the LO expression for $T_\Gamma^{\rm I}$ following from (\ref{me1}) we find
that only two independent functions of the quark momentum fraction $u$ appear, 
which can be taken as\footnote{With this choice we obtain simple expressions
in the limit $k^2 \to 2 E_1 M_B$, namely $f_1(u) \to E_1/E_2$ and $f_2(u) \to 1/\bar u$.}
\begin{align}
f_1(u) &\equiv \frac{-k^2}{\bar u \, (k^2-2 E_1 M_B)-2 E_2 M_B} \,,
\qquad 
f_2(u) \equiv \frac{2 E_2 M_B}{\bar u \, k^2} \, f_1(u) \,.
\label{f12def}
\end{align}
The moment $\langle \bar u^{-1}\rangle_\pi$ can be obtained from a linear combination,
\begin{align}
\frac{1}{\bar u} = \left( \frac{2 E_1 M_B}{k^2}-1 \right) f_1(u)  + f_2(u) \,.
\end{align}
With these definitions we obtain
\begin{align}
T_\Gamma^{\rm I}(u,k^2,E_1,E_2) \Big|_{\rm LO}
&=  i \,
\frac{\alpha_s C_F}{N_C} 
\left\{ 
f_1(u) \left[ \left( \frac{2 E_1 M_B}{k^2}-1 \right) s_2 + \frac12 \, s_3 \right] 
\right. 
\cr 
& \phantom{ i \,
\frac{\alpha_s C_F}{N_C} } \left. \qquad {} + 
f_2(u) \left[ s_1 + s_2 - \frac{M_B}{2 E_2} \, s_5 - \frac12 \, s_7 \right]
\right\}
\cr 
&\equiv i \,
\frac{\alpha_s C_F}{N_C} \, \frac{S_A + S_B^{(i)}(u)+ S_B^{(ii)}(u)}{\bar u}\,,
\label{T1exp}
\end{align}
where, for later use, we have defined the abbreviations
\begin{align}
S_A &=   s_2
\,, \ \quad 
\frac{S_B^{\rm (i)}(u)}{\bar u} = \frac{f_1(u)}{2} \, s_3 - \frac{M_B \, f_2(u)}{2 E_2} \, s_5
\,, \ \quad  
\frac{S_B^{\rm (ii)}(u)}{\bar u} = f_2(u) \left[ s_1 - \frac{s_7}{2} \right] 
\,.
\label{T1}
\end{align}

Notice that in the individual contributions to $T_\Gamma^{\rm I}$,
different projections of the Dirac matrix $\Gamma$ in the original $b \to u$ transition
current appear. In particular, at LO, the hard-gluon exchange involves the ``small'' spinor
components, $(1-P_v) \, \psi_b$ for the heavy quark (in the Dirac structures $s_{3,7}$), and 
$\frac{\slashed k_1 \slashed k_2}{k^2} \, \psi_u$ for the emitted $u$-quark (in the Dirac structures $s_{2,6}$), 
but not both of them simultaneously (i.e.\ the structures $s_4$ and $s_8$ do not appear).

\subsubsection{Twist-3 contributions}

As is known from the QCDF analysis of $B \to \pi\pi$ decays \cite{Beneke:2001ev},
twist-3 contributions to the hard-scattering kernels can be numerically important,
despite the fact that they are formally power-suppressed. This can be traced back
to a large numerical pre-factor, $\mu_\pi=m_\pi^2/(m_u+m_d) \sim 2.5~$GeV, which is proportional
to the quark condensate in QCD. The power corrections of the order $\mu_\pi/\sqrt{k^2}$
will therefore be refered to as {\it chirally enhanced}. In addition, power corrections
will potentially lead to non-factorizable contributions which show up as 
endpoint-divergent integrals in the perturbative calculation.
In the computation of the kernel $T_\Gamma^{\rm I}$ the chirally-enhanced terms
arise from the twist-3 two-particle LCDAs of the $\pi^+$ meson.
Here, a comment is in order about the definition of the transverse plane related to the 
underlying light-cone expansion for the \emph{positively} charged pion state:
As can be seen from the explicit structure of the LO diagrams leading to 
(\ref{bpid}), the gluon propagator associated to the separation of the quark fields in 
the $|\pi^+\rangle$ state involves the large momenta $(p_b-q)^\mu \simeq (k_1^\mu+k_2^\mu)$
and $k_1^\mu$. The transverse momenta in the light-cone expansion for the $\pi^+$
matrix elements are therefore to be chosen as transverse to 
both pion momenta, $k_1$ and $k_2$. The parton momenta in the two-particle
Fock state are then expanded as
\begin{align*}
\mbox{up-quark in $\pi^+$:} \qquad & k_{q1}^\mu \simeq u k_1^\mu + k_\perp^\mu \,,
\cr 
\mbox{anti-down-quark in $\pi^+$:} \qquad & k_{\bar q1}^\mu \simeq \bar u k_1^\mu - k_\perp^\mu \,,
\qquad \mbox{with $k_{1,2} \cdot k_\perp \equiv 0$} \,,
\label{twist31}
\end{align*}
with $|k_\perp|$ scaling as a hadronic momentum of order $\Lambda$.
The corresponding
twist-3 momentum-space projector can then be written as (see also \cite{Beneke:2000wa})
\begin{align}
 {\cal M}_{\pi^+}^{(3)}(u) &= 
 \frac{i f_\pi \mu_\pi}{4} \, \frac{\mathds{1}}{N_C} \, \gamma_5
 \left\{ 
   -  \phi_P(u)
   + i  \sigma_{\mu\nu} \, \frac{k_1^\mu k_2^\nu}{k_1 \cdot k_2} \,
    \frac{\phi'_\sigma(u)}{6} 
    - i \sigma_{\mu\nu} \, \frac{\phi_\sigma(u)}{6} \, k_1^\mu \frac{\partial}{\partial k_{\perp\nu}}
 \right\} \Big|_{k_\perp \to 0} \,.
\end{align}
Neglecting 3-particle contributions, the corresponding LCDAs are fixed by the equations
of motion (see e.g.\ \cite{Braun:1989iv}),
\begin{align}
 &  \frac{u}{2} \left( \phi_P(u) + \frac{\phi_\sigma'(u)}{6} \right) \simeq 
    \frac{\bar u}{2} \left( \phi_P(u) -\frac{\phi_\sigma'(u)}{6} \right) \simeq \frac{\phi_\sigma(u)}{6} \,,
\end{align}
leading to
\begin{align}
 \phi_P(u)\simeq 1 \,, \qquad  \phi_\sigma(u) \simeq 6 u \bar u \,. \qquad  \mbox{(``Wandzura-Wilczek approx.'')}
 \label{eom1}
\end{align}
The twist-3 analogue to (\ref{bpid}) can then be derived from
\begin{align}
\Gamma_X & \to  - \frac{\gamma_\alpha  \, {\cal M}_{\pi^+}^{(3)}(u) \, \gamma^\alpha  \,
 (\slashed p_b -\slashed q) \, \Gamma}{(p_b-q)^2 \, (p_b - q -  u k_1)^2} 
- \frac{\gamma_\alpha  \, {\cal M}_{\pi^+}^{(3)}(u) \, \Gamma \, (u \slashed k_1+ \slashed k_\perp 
  +\slashed q + m_b)
 \, \gamma^\alpha  }{[(u k_1 + q)^2 + 2 \, k_\perp \cdot q-m_b^2]\, (p_b - q - u k_1)^2} \,.
\label{bpid3}
\end{align}
The corresponding contributions to the $B \to \pi\pi$ matrix elements can be written as 
\begin{align}
 &\langle \pi^+(k_1)\pi^-(k_2)|\bar \psi_u \, \Gamma \, \psi_b |B^-(p)\rangle \Big|_{\text{twist-3, LO}}
\cr 
&= \frac{2\pi \, f_\pi}{k^2} \, \xi_{\pi}(E_2;\mu) \,
 \int_0^1 du \, \left( \phi_P(u) \, T_\Gamma^{(\rm I,P)}(u,k^2,E_1,E_2) + \phi_\sigma(u) \, T_\Gamma^{(\rm I,\sigma)}(u,k^2,E_1,E_2) \right)
 \,.
 \cr & 
\end{align}
(Notice that -- from the approximate 
relations in (\ref{eom1}) --  there is an ambiguity
in expressing $\phi_\sigma'(u)$ in terms of $\phi_\sigma(u)$ and $\phi_P(u)$.)
The first term in (\ref{bpid3}) contributes
\begin{align}
  T_\Gamma^{(\rm I,P)} &=  i \,
\frac{\alpha_s C_F}{N_C} \, \frac{2 M_B \mu_\pi}{k^2} \, \frac{s_5}{\bar u} \,.
\label{TP}
\end{align}
The second term in (\ref{bpid3}) contributes
\begin{align}
  T_\Gamma^{(\rm I,\sigma)} &=  i \,
\frac{\alpha_s C_F}{N_C} \, \frac{ M_B \mu_\pi}{3 \left( \bar u (k^2- 2 M_B E_1)-2 M_B E_2\right)} 
\cr & \qquad \times 
\left\{ 
  \frac{1}{\bar u} \left[ - s_2 - \frac{E_2}{M_B} \, s_4 + \frac{2 E_2 M_B}{k^2} \, s_6 + \frac{s_7}{2} \right] 
  \right. 
  \cr 
  & \phantom{\times  }
  \qquad \left. 
  + \frac{1}{u} \left[ - s_2 - \frac{E_2}{M_B} \, s_4 + \frac{2 E_2 M_B}{k^2} \, s_6 + \frac{s_8}{2}  \right] 
  + \frac{2 E_2 M_B}{k^2} \, \frac{s_5}{\bar u^2} 
\right\}
 \cr 
 & \ + i \,
\frac{\alpha_s C_F}{N_C} \, \frac{2 E_2 M_B^2 \mu_\pi}{3 \left( \bar u (k^2- 2 M_B E_1)-2 M_B E_2\right)^2} 
\cr & \qquad \times 
\left\{ 
  \frac{1}{\bar u} \left[ 
  \frac{E_2}{M_B} \, s_3 -  s_5 + \frac{s_7}{2} + \frac{ (4 E_1 E_2 - k^2) \, M_B}{2 E_2 k^2} \, s_5 
  \right] 
  \right. 
  \cr 
  & \phantom{\times  \frac{1}{\bar u} }
  \qquad \left. 
  - \left[ 
   \frac{k^2}{2 E_2 M_B} \left(  s_1 - \frac{s_3}{2} \right) - \frac{E_1}{E_2} \, \frac{s_7}{2}
  \right] 
\right\}
 \,,
\label{Tsig}
\end{align}
where we have used the Wandzura-Wilczek approximation in (\ref{eom1}).
Notice that the potential endpoint divergence from the term $\phi_P(u)/\bar u$ in the limit $\bar u \to 0$ 
in (\ref{TP}) cancels with the last term in the first curly brackets in (\ref{Tsig}).
This does not necessarily need to remain true after spectator-scattering
corrections are taken into account, i.e.\ the contributions to the
kernel $T_\Gamma^{\rm II}$ involving the twist-3 LCDAs of the positively charged
pion can be expected to exhibit additional endpoint-divergent expressions, similar
to what is observed in the QCDF approach to non-leptonic $B \to \pi\pi$ decays.
In the  approximation (\ref{eom1}) the convolution integrals with respect to the quark momentum fraction $u$
can be done explicitly, leading to
\begin{align}
 & \int_0^1 du \, \left( \phi_P(u) \, T_\Gamma^{(\rm I,P)}(u,k^2,E_1,E_2) + \phi_\sigma(u) \, T_\Gamma^{(\rm I,\sigma)}(u,k^2,E_1,E_2) \right)
 \cr 
 &\simeq \frac{2 M_B \mu_\pi}{k^2}
 \left( (1+L) \, s_5 - \frac{E_2}{E_1} \, L \, s_6 \right)
 \cr 
 & \quad - \frac{2 M_B \mu_\pi \, L}{k^2 - 2 M_B E_1} 
 \left( s_2 + \frac{E_2}{M_B} \, s_4 - \frac{E_2}{E_1} \, s_6 \right)
 \cr 
 & \quad - \frac{2 M_B \mu_\pi}{k^2 - 2 M_B E_1} 
 \left[1 + \frac{2 M_B E_2}{k^2 - 2 M_B E_1} \, L \right]
 \left( \frac{E_2}{M_B} \, s_3 - \frac{M_B}{2E_2} \, s_5 - \frac{s_8}{2} 
 \right)
 \cr 
 & \quad +
 \frac{2 M_B k^2 \mu_\pi}{(k^2-2 M_B E_1)^2} 
 \left[ 1 + \left( \frac{2 M_B E_2}{k^2 - 2 M_B E_1} - \frac12 \right)
  L \right] 
 \left( 2 s_1 - s_3 - s_7 \right) 
 \label{twist3contr}
\end{align}
with
\begin{align}
  L &\equiv \ln\left[\frac{2 M_B E_1 + 2 M_B E_2 - k^2}{2 M_B E_2} \right]
  = \ln \left[\frac{M_B^2-q^2}{2 M_B E_2} \right] \,.
\end{align}
Notice that the twist-3 contributions to $T_\Gamma^{\rm I}$ 
now also involve the Dirac structures $s_{4,6,8}$ which 
did not appear in (\ref{T1}).

\subsection{The kernel $T^{\rm II}$}

\begin{figure}[t]
\begin{center}
\includegraphics[width=0.98\textwidth]{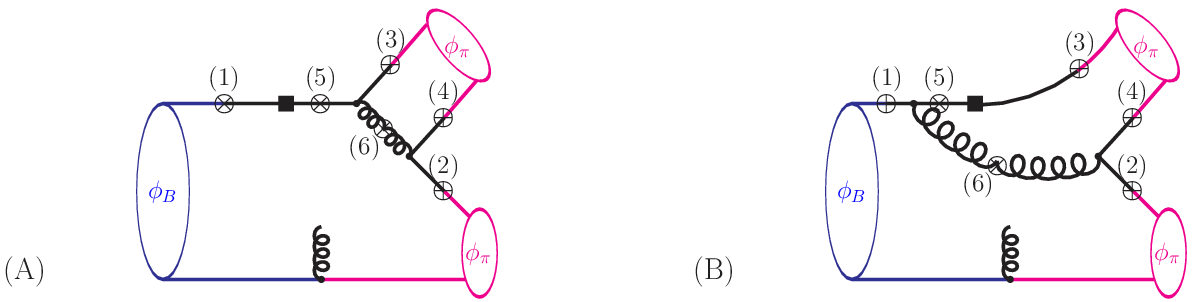}
\end{center}
\caption{\label{fig2}
Diagrams contributing at LO to the kernel $T^{\rm II}$. The hard-collinear
gluon emitted from the lower quark line can be connected to any of the 
crosses numbered by $(1-6)$.}
\end{figure}

The leading contribution to the kernel $T^{\rm II}$ in the QCD factorization
formula (\ref{theorem}) arises from diagrams where -- in addition to 
the hard-gluon process in Fig.~\ref{fig1} -- a ``hard-collinear''
gluon connects to the (would-be) spectator quark in the $B$-meson. 
The relevant Feynman diagrams are summarized in Fig.~\ref{fig2},
and will be discussed in turn in Appendix~\ref{app:t2}.

Again, a comment is in order about the definition of the transverse plane, now related to the 
underlying light-cone expansion for the \emph{negatively} charged pion state:
In contrast to the situation discussed around (\ref{twist31}) for the partonic kinematics in 
the $|\pi^+\rangle$ state, 
the hard-collinear gluon propagator associated to the separation of the quark fields in 
the $|\pi^-\rangle$ state involves the large momenta $p_b^\mu \sim p^\mu$
and $k_2^\mu$. The transverse momenta in the light-cone expansion for the $\pi^-$
matrix elements are therefore conveniently chosen as transverse to 
$p$ and $k_2$. The parton momenta in the two-particle
Fock state are then expanded as
\begin{align*}
\mbox{down-quark in $\pi^-$:} \qquad & k_{q2}^\mu \simeq v k_2^\mu + \bar k_\perp^\mu \,,
\cr 
\mbox{anti-up-quark in $\pi^-$:} \qquad & k_{\bar q2}^\mu \simeq \bar v k_2^\mu - \bar k_\perp^\mu \,,
\qquad \mbox{with $k_2 \cdot \bar k_\perp = p \cdot \bar k_\perp \equiv 0$} \,,
\end{align*}
with $v$ ($\bar v = 1-v$) denoting the longitudinal momentum fraction of the 
quark (anti-quark), and $|\bar k_\perp|$ scaling as a hadronic momentum
of order $\Lambda$.
The corresponding twist-3 momentum-space projector should then be written as 
\begin{align}
 {\cal M}_{\pi^-}^{(3)}(v) &= 
 \frac{i f_\pi \mu_\pi}{4} \, \frac{\mathds{1}}{N_C} \, \gamma_5
 \left\{ 
   -  \phi_P(v)
   + i  \sigma_{\mu\nu} \, \frac{k_2^\mu p^\nu}{p \cdot k_2} \,
    \frac{\phi'_\sigma(v)}{6} 
    - i \sigma_{\mu\nu} \, \frac{\phi_\sigma(v)}{6} \, k_2^\mu \frac{\partial}{\partial \bar k_{\perp\nu}}
 \right\} \Big|_{\bar k_\perp \to 0} \,.
\end{align}
Neglecting 3-particle contributions, the corresponding LCDAs 
will again be fixed by the equations 
of motion as in (\ref{eom1}).

With the same argument, we define the transverse momenta $l_\perp$ 
of the light anti-quark in the $B$-meson, such that the 
momentum-space projector for the 2-particle distribution amplitudes
can be written as in \cite{Beneke:2000wa},
\begin{align}
 {\cal M}_{B}^{(WW)}(\omega) &= 
 - \frac{i f_B M_B}{4} \, \frac{\mathds{1}}{N_C} \left[
 P_v \left\{ \phi_B^+(\omega) \, \slashed n_+ 
  + \phi_B^-(\omega) \left( \slashed n_- - \omega \gamma_\perp^\nu 
   \frac{\partial}{\partial l_\perp^\nu} \right) \right\} \gamma_5 
   \right]_{l_\perp \to 0} \,,
\end{align}
where $v_b^\mu=p^\mu/M_B$, $n_-^\mu = k_2^\mu/(v_b \cdot k_2)$ and $ n_+^\mu =
2 v_b^\mu - n_-^\mu$, and $\omega = (n_- \cdot l)$ is the light-cone projection
of the light anti-quark momentum. As indicated, we again work in the
Wandzura-Wilczek approximation and neglect the 3-particle DAs.

The individual  contributions from a given diagram X to the $B \to \pi\pi$ matrix element 
will be decomposed 
as follows,
\begin{align}
 &\langle \pi^+(k_1)\pi^-(k_2)|\bar \psi_u \, \Gamma \, \psi_b |B^-(p)\rangle \Big|_{\rm (Diagram X)}
\cr 
&= \frac{2\pi f_\pi}{k^2} \, 
\frac{i\alpha_s^2 C_F}{4\pi N_C} \, \frac{\pi^2 f_B f_\pi M_B}{N_C E_2^2}
\, \int_0^1 du \, \phi_\pi(u) \,  \int_0^1 dv \, \int_0^\infty d\omega
\left( g_{(X)}^{\rm finite} + g_{(X)}^{\rm endpoint} \right) \,.
\label{gdecomp}
\end{align}
Detailed inspection of the diagrams in Fig.~\ref{fig2} 
reveals that the corresponding contributions can be calculated 
in a similar way as the spectator-scattering contributions to
the $B \to \pi$ form factors considered in  \cite{Beneke:2000wa}
at leading non-vanishing order. In particular, we find that all
the endpoint-sensitive (formally divergent) contributions from 
2-particle Fock states at leading power in the $1/M_B$ expansion 
can be absorbed into the universal form factor $\xi_\pi$, with 
the definition of the associated hard kernel $T_\Gamma^{\rm I}$
derived in Eq.~(\ref{T1}). The details of the calculation for the 
individual subdiagrams can be found in Appendix~\ref{app:t2}.

\subsubsection{Endpoint-divergent terms}

\begin{table}[ptb]
\begin{center} \renewcommand{\arraystretch}{1.8}
\begin{tabular}{|c || c |  c | c | c| c|| c|
}
\hline 
structure & A1 & A2 & A3 + A4 & A5 & A6 & A1-A6
\\
\hline \hline 
$\frac{2 E_2 M_B}{\bar u^2  k^2}\, s_5 \,
 \frac{\phi_B^+(\omega)}{\omega} \,
 \frac{\phi_\pi(v)}{2 v \bar v}$
 & 0 & 0 & $ - C_{FA} \, 2v $ & 0 & $C_A \, \frac{v-\bar v}{2}$ 
 & $2 v C_F - \frac{C_A}{2}$
 \\
 \hline 
 $\frac{S_A}{\bar u} 
\, \frac{\phi_B^-(\omega)}{\omega} \,
\, \frac{\phi_\pi(v)}{\bar v^2}$
& $C_F \, \frac{1}{v}$ & $ C_F \, \bar v$ & $C_{FA} \, \frac{\bar v}{v}$  
& 0 
& $-\frac{C_A}{2} \, \frac{\bar v}{v}$ 
& $C_F \, (1+\bar v)$
\\
\hline 
$ \frac{S_A}{\bar u} 
\, \frac{\phi_B^+(\omega)}{\omega} \, 
 \frac{\mu_\pi \phi_\sigma(v)}{6\bar v^3 E_2}$
& $C_F$
& 0
& 0 
& 0
& 0
& $C_F$
\\
\hline 
 $ 2 \mu_\pi \, \frac{S_A}{\bar u} 
\,\frac{\phi_B^+(\omega)}{\omega^2} \, 
 \frac{\phi_P(v)}{\bar v}$
 & 0
 & $C_F$
 & 0
 & 0
 & 0
  & $C_F$
 \\
 \hline 
\multicolumn{7}{c}{}
\\
 \hline 
structure & B1 & B2 & B3+B5 & B4 & B6 & B1-B6
\\
\hline \hline 
$
\frac{2 E_2 M_B}{\bar u^2  k^2}\, s_5 \, 
 \frac{\phi_B^+(\omega)}{\omega} \,
 \frac{\phi_\pi(v)}{2v \bar v}$
 & 0 & 0 & 0 
 & $  C_{FA} \, 2v $ 
 & $ C_A \, \frac{\bar v -v}{2}$ 
 & $\frac{C_A}{2}- 2 v C_F$
 \\
 \hline 
 $
\frac{S_B^{\rm (i)}}{\bar u} \, 
\frac{\phi_B^+(\omega)}{\omega} \, 
 \frac{\phi_\pi(v)}{\bar v^2}
$
 & 0 & 0 & $ - C_{FA} \, v_\perp^2 $ & $ C_{FA} \, v_\perp^2$ & 0 & 0 
 \\
 \hline
 $\frac{S_B^{\rm (i)}+S_B^{\rm (ii)}}{\bar u} 
\, \frac{\phi_B^-(\omega)}{\omega} \,
\, \frac{\phi_\pi(v)}{\bar v^2}$
& $C_F \, \frac{1}{v}$ & $ C_F \, \bar v$ & $C_{FA} \, \frac{1}{v}$ & $-C_{FA}$ 
& $-\frac{C_A}{2} \, \frac{\bar v}{v}$ 
& $C_F \, (1+\bar v)$
\\
\hline 
$ \frac{S_B^{\rm(i)} + S_B^{\rm (ii)}}{\bar u} 
\, \frac{\phi_B^+(\omega)}{\omega} \, 
 \frac{\mu_\pi \phi_\sigma(v)}{6\bar v^3 E_2}$
& $C_F$
& 0
& $- C_{FA} \, v_\perp^2 $
& $ C_{FA} \, v_\perp^2$
& 0
& $C_F$
\\
\hline 
$ \frac{S_B^{\rm (i)}}{\bar u}
\,\frac{\phi_B^-(\omega)}{\omega} \, 
 \frac{\mu_\pi \phi_\sigma(v)}{6\bar v^3 E_2}$
 & 0 
 & 0
 & $C_{FA} \, v_\perp^2$
 & $-C_{FA} \, v_\perp^2$
 & 0
  &0
 \\
 \hline
 $ 2 \mu_\pi \, \frac{S_B^{(\rm i)}+S_B^{\rm (ii)}}{\bar u} 
\,\frac{\phi_B^+(\omega)}{\omega^2} \, 
 \frac{\phi_P(v)}{\bar v}$
 & 0
 & $C_F$
 & 0
 & 0
 & 0
  & $C_F$
 \\
 \hline
\end{tabular}
\end{center}
\caption{\label{tab:endpoint} Endpoint-divergent contributions $g_{(X)}^{\rm endpoint}$
from diagrams (A1-A6) and (B1-B6) in Feyman gauge.}
\end{table}

In Table~\ref{tab:endpoint}, we summarize the results for the endpoint-divergent
terms as appearing in the individual diagrams when calculated in
Feynman gauge. 
Here, we have introduced the additional abbreviations
\begin{align}
-v_\perp^2 &= \frac{4E_1 E_2}{k^2} - 1 \,,
\label{vperpdef}
\end{align}
where $v_\perp^\mu$ denotes the transverse components of the $b$-quark velocity
with respect to the $k_1$--$k_2$ plane,
and 
\begin{align}
  C_{FA} & = \frac{C_A}{2} - C_F = \frac{1}{2N_C} \,,
  \label{CFAdef}
\end{align}
for the coefficient of the sub-leading colour structure.
We further use Eq.~(\ref{eom1}) to replace
\begin{align}
\frac{\mu_\pi}{2 E_2} \left( \phi_P(v) - \frac{\phi_\sigma'(v)}{6}\right) 
 & \simeq \frac{\mu_\pi \phi_\sigma(v)}{6 \bar v E_2} \,. 
 \label{phi3effdef}
\end{align}
We observed that some obvious 
cancellations (of sometimes rather complicated structures)
appear inbetween diagrams (A3,A4) and (B3,B5), respectively. 
For the sake of readability, we only show the combined results.
The final expression for the endpoint-divergent terms
arises as the result of rather non-trivial cancellations
among the individual diagrams, see Table~\ref{tab:endpoint}.
This also involves the cancellation of 
endpoint-divergences related to the momentum
fraction $\bar u\to 0$ of the anti-quark in the \emph{positively} charged pion, 
as expected from colour-transparency arguments \cite{Beneke:1999br}.
We obtain
\begin{align}
 &\langle \pi^+(k_1)\pi^-(k_2)|\bar \psi_u \, \Gamma \, \psi_b |B^-(p)\rangle \Big|_{\rm (A1-A6,\,B1-B6)}
\cr  
&= \frac{2\pi f_\pi \, \xi_\pi^{\rm (HSA)}(E_2)}{k^2} \, 
\int_0^1 du \, \phi_\pi(u) \,  T_\Gamma^{\rm I} (u,k^2,E_1,E_2)
+ \mbox{finite terms,}
\label{A16B16}
\end{align}
where the corresponding endpoint-divergent contributions in
$\xi_\pi^{\rm (HSA)}(E_2)$ have been calculated in \cite{Beneke:2000wa}
and can be found in Eq.~(\ref{xihsa}) in the appendix.
We thus recover the very same structures as in (\ref{T1exp}), 
confirming the assumptions that we made in the derivation of $T_\Gamma^{\rm I}$ in Section~\ref{sec:T1}. 
Notice that in Feynman gauge all diagrams (except for A5) contribute, and the correct cancellation/combination of endpoint-divergences provides
a useful cross-check of our calculation and a non-trivial aspect for the confirmation
of the factorization hypothesis.

\subsubsection{Finite Terms}

The remaining (endpoint-finite) terms can then be associated 
to the kernel $T_\Gamma^{\rm II}$, thus verifying the factorization
formula (\ref{theorem}) to leading order in the perturbative expansion.

\vspace{1em}

\paragraph{Large-$\boldmath N_C$ limit:}
Neglecting corrections that vanish in the limit $N_C \to \infty$ (which 
amounts to setting $C_A = 2 C_F$),
the hadronic information in the LO expression for $T_\Gamma^{\rm II}$
can be encoded in terms of the functions
\begin{align}
 f_3(u,v) &= \frac{\phi_\pi(v)}{\bar u \, v } \,, \qquad f_4(u,v) = \frac{\phi_\pi(v)}{\bar u \, v \, \bar v} \, ,
 \cr 
 f_5(u,v) &= 
 \frac{4 v E_2 \, (k^2-E_1 M_B) + \bar v \, k^2 M_B}{v \bar v \, k^2 \, M_B} \, f_1(u) \,, 
 \cr 
 f_6(u,v) &= 
 \frac{4 v E_2 \, (k^2-E_1 M_B) + \bar v \, k^2 M_B}{v \bar v \, k^2 \, M_B} \, f_2(u)
 \,.
\end{align}
Notice that only three of these functions are linearly independent, since 
\begin{align}
& f_6(u,v) + \left( \frac{2 E_1 M_B}{k^2} - 1 \right) f_5(u,v) 
 \cr 
& \quad 
 + \left( \frac{4 E_1 E_2}{k^2} - \frac{4 E_2}{M_B} \right) f_4(u,v) 
 - \left( 1 - \frac{4 E_2}{M_B} + \frac{4 E_1 E_2}{k^2} \right) f_3(u,v)  = 0 \,.
 \label{3456rel}
\end{align}
The explicit computation of the individual diagrams 
in Feynman gauge (see appendix~\ref{app:t2}) yields
\begin{align}
 g_{(A1-A6)}^{\rm finite} \Big|_{C_A = 2 C_F}
 &= C_F \left\{ f_3(u,v) \left( s_2 - \frac{2 E_2 M_B}{k^2} \, s_6 \right)
 \right. 
 \cr 
 & \phantom{C_F} \qquad \left. 
 + f_4(u,v) \left( \frac{E_2}{M_B} \, s_4 + \frac{2 E_2 M_B}{k^2} \, s_5 \right)
 \right\} \frac{\phi_B^{+}(\omega)}{\omega} \,,
 \label{gA16a}
\end{align}
and 
\begin{align}
 g_{(B1-B6)}^{\rm finite}\Big|_{C_A = 2 C_F} &= C_F 
\left\{ 
- f_3(u,v) \, \frac{2 E_2}{M_B} \, s_3
+
f_4(u,v) \, \frac{E_2}{M_B} \, s_3 
\right. 
\cr 
& \phantom{C_F} \qquad \left.
+
f_5(u,v) \, \frac{s_3}{2} 
-
f_6(u,v) \, \frac{M_B}{2 E_2} \, s_5
\right\}
\frac{\phi_B^{+}(\omega)}{\omega} \,.
 \label{gB16a}
\end{align}

As a consequence of (\ref{3456rel}), the results only depends on \emph{three} new
independent Dirac structures, which can be chosen as 
\begin{align*}
& \left[ s_2 + \frac{ M_B (4 E_1 E_2 - k^2)}{2 E_2 k^2} \, s_5 - \frac{2 E_2 M_B}{k^2} \, s_6 \right]
 \,, \cr 
 & \left[ s_3 - \frac{M_B (k^2-2 E_1 M_B)}{E_2 k^2} \, s_5 \right] \,, \qquad 
  \left[ s_4 - \frac{M_B (k^2-2 E_2 M_B)}{E_2 k^2} \, s_5 \right] \,.
\end{align*}

\vspace{1em}

\paragraph{Subleading terms in $\boldmath 1/N_C$:}

Including finite terms of order $(\frac{C_A}{2}-C_F)=\frac{1}{2N_C}$, which arise from 
the diagrams $B_3$ and $B_5$, we encounter two more hadronic functions,
\begin{align}
f_7(u,v) &\equiv 
\frac{-2 E_2 M_B}{\bar u (v k^2 - 2  E_1 M_B)- 2 v E_2 M_B} \, f_4(u,v) \,,
\cr 
f_8(u,v) & \equiv \frac{ \bar u k^2 \, (M_B-2  v \, E_2 )  + 4 v \, E_2^2 M_B}{2 E_2 \,
(\bar u (k^2 - 2 E_1 M_B)-2 E_2 M_B))} f_7(u,v) \,,
\end{align}
entering as 
\begin{align}
 g_{(B1-B6)}^{\rm finite}\Big|_{\frac{C_A}{2}-C_F} &= \left( \frac{C_A}{2} - C_F \right) 
\left\{ 
- \left(f_7(u,v) + f_8(u,v)\right)  \frac{E_2}{M_B} 
\left[ s_3  - \frac{M_B (k^2-2 E_1 M_B)}{E_2 k^2} \, s_5 \right]
\right. 
\cr 
& \phantom{\left( \frac{C_A}{2} - C_F \right) } \qquad \left.
- 
f_7(u,v) \left[ \frac{s_7}{2} - \frac{M_B}{2 E_2} \, s_5 \right]
\right\}
\frac{\phi_B^{+}(\omega)}{\omega} \,.
 \label{gB16b}
\end{align}
This involves another independent 
Dirac structure, $\left[ \frac{s_7}{2} - \frac{M_B}{2 E_2} \, s_5 \right]$.

\vspace{1em}

\paragraph{Final result for $T_\Gamma^{\rm II}$:}

For the very definition of $T_\Gamma^{\rm II}$, we have to specify 
the factorization prescription for the soft form factor $\xi_\pi(E_2)$.
If we identify $\xi_\pi(E_2)$ with the physical form factor 
$f_+((p-k_2)^2)$ for $B\to\pi$ vector transitions, with $(p-k_2)^2=M_B^2-2 M_B E_2$,
we obtain 
\begin{align}
 & \phi_\pi(v) \, \frac{\phi_B^+(\omega)}{\omega} \, T_\Gamma^{\rm II}(u,v,\omega,k^2,E_1,E_2) 
 \cr 
 &= 
 g_{(A1-A6)}^{\rm finite} + g_{(B1-B6)}^{\rm finite} - g_+^{\rm finite}(v,\omega,E_2) \, T_\Gamma^{\rm I}(u,k^2,E_1,E_2) \,.
\end{align}
Here the function $g_{(A1-A6)}^{\rm finite}$ and 
$g_{(B1-B6)}^{\rm finite}$ can be found in Eqs.~(\ref{gA16a}), (\ref{gB16a}), (\ref{gB16b}), and 
the finite contributions to the $B \to \pi$ form factor $f_+(E_2)$ 
are encoded in the function $g_+^{\rm finite}$ as given in Eq.~(\ref{gplus}) in
the appendix.

\section{$B \to \pi\pi$ Form Factors and Observables}

\label{sec:obs}

We are now going to briefly discuss some general phenomenological
implications of the factorization formula (\ref{theorem}) for 
the $B \to \pi\pi$ form factors and $B^- \to \pi^+\pi^-\ell^-\bar\nu_\ell$
decay observables in the kinematic region of small momentum
transfer $q^2$ and large dipion mass $k^2$.

\subsection{Reduction of independent form factors in the QCDF limit}

We first observe that the leading-twist contribution to the 
LO expression for the 
kernel $T_\Gamma^{\rm I}$ involves only two independent 
Dirac structures, see Eq.~(\ref{T1exp}).
Introducing
\begin{align}
    S_1(\Gamma) & \equiv \left(\frac{2 E_1 M_B}{k^2} - 1\right) s_2 +
\frac{1}{2} s_3\,, &
    S_2(\Gamma) & \equiv s_1 + s_2 - \frac{M_B}{2 E_2} s_5 - \frac{1}{2}
s_7\,,
\end{align}
we thus have
\begin{align}
&    \langle \pi^+(k_1)\pi^-(k_2)|\bar \psi_u \, \Gamma \, \psi_b
|B^-(p)\rangle\Big|_{\rm twist-2} \cr 
 &      \simeq\frac{2\pi f_\pi}{k^2} 
        \left\{ S_1(\Gamma) \, F_1(k^2,q^2,q\cdot \bar k) +
S_2(\Gamma) \, F_2(k^2,q^2,q\cdot \bar k) \right\} \,,
\label{F12expr}
\end{align}
up to higher-order corrections in the strong coupling.
The form factors $F_{1,2}(k^2,q^2,q\cdot \bar k)$ follow from the LO expression 
for the kernel $T_\Gamma^{\rm I}$ in (\ref{T1exp}),
\begin{equation}
    F_{1,2}(k^2,q^2,q\cdot \bar k) 
    \equiv  \xi_\pi(E_2,\mu) \, \frac{i\alpha_s(\mu) \, C_F}{N_C} \,
\int_0^1 du\,\phi_\pi(u,\mu) \, f_{1,2}(u)\,,
\end{equation}
where the functions $f_{1,2}(u)$ are defined in Eq.~(\ref{f12def}),
and the dependence on the kinematic variables follows from
Eq.~(\ref{kinrel}).

As explained above, the twist-3 contributions 
in (\ref{twist3contr}) are formally power-suppressed, but numerically
of the same order as the twist-2 terms because 
$\mu_\pi/E_1 \simeq {\cal O}(1)$, and therefore they 
have to be included as well. 
On the other hand, the spectator interactions contributing 
to the kernel $T_\Gamma^{\rm II}$ are suppressed by the strong
coupling constant and can be neglected to first approximation.

\subsubsection{Relations among partial-wave form factors}

\begin{table}[t!pbh]
\begin{center}
 \begin{tabular}{|c|ccc|}
   \hline 
   & $S$-wave & $P$-wave & $D$-wave 
   \\ 
   \hline \hline 
   $F_0$ & $\sqrt\lambda$ & $1$ & $\sqrt\lambda$ 
   \\
   $F_t$ & $1$ & $\sqrt\lambda$ & $\lambda$
   \\
   $F_\perp$ & -- & $\sqrt\lambda$ & $\lambda$ 
   \\
   $F_\parallel$ &-- & $ 1$ & $\sqrt\lambda$ 
   \\
   \hline 
 \end{tabular}
\caption{\label{tab:pwscale} Scaling of partial-wave form factors 
as defined in Appendix~\ref{app:ff} with $\sqrt\lambda$.}
\end{center}

\end{table}

From (\ref{me1}) and (\ref{T1}) and (\ref{twist3contr})
we can easily compute the leading contributions to vector and axial-vector form factors.
To this end, we first project onto helicity form factors as defined in \cite{Faller:2013dwa}
and summarized in Eq.~(\ref{A5}) in the appendix.
Using that for the phase space Scenarios A and B $$q^2 \sim \sqrt{\lambda} \ll M_B^2 \,,$$
each helicity form factor can then be expanded in the small parameter 
$\Delta E_{\pi}/M_B \sim \sqrt{\lambda}/M_B^2$ which, via (\ref{kin2}), translates into a power series in 
the angular variable $z\equiv\cos\theta_\pi$. From this, it is a straightforward task to 
identify the leading contributions to particular partial waves where 
-- as a general rule, with one exception,\footnote{Notice that -- in the considered kinematic 
region -- the $S$-wave contribution
to the form factor $F_0$ is suppressed compared to the $P$-wave and of the same order
as the $D$-wave. This differs from other kinematic situations as considered 
e.g.\ in \cite{Faller:2013dwa}. In particular, the form factor $F_0^{(D)}$ will now also 
provide a leading contribution  
to the forward-backward asymmetry with respect to the polar angle $\theta_\pi$.
} see Table~\ref{tab:pwscale}
-- higher partial waves will be suppressed by increasing
powers of $\sqrt{\lambda}/M_B$. 
Performing the Gegenbauer expansion of the twist-2 pion LCDA to 
second order, the leading twist-2 and twist-3 contributions
to the partial-wave form factors are obtained as  
\begin{align}
 F_0^{(S)}  \approx \frac{\sqrt\lambda}{2 M_B \sqrt{q^2}}\, F_t^{(S)}
 &\ \approx \ 
 \frac{i \alpha_s C_F}{N_C} \, \frac{2 \pi f_\pi}{M_B} \, \frac{2\sqrt{\lambda}}{M_B \sqrt{q^2}}
  \left( 1+\frac{3  a_2^\pi}{4} + \frac{\mu_\pi}{M_B} \right) \xi_\pi(\frac{M_B}{2}) \,, 
  \label{Swaverelation}
\end{align}
and 
\begin{align}
  F_0^{(P)} \simeq \frac{1}{\sqrt 2} \, F_\parallel^{(P)} \approx \frac{2 M_B \sqrt{q^2}}{\sqrt\lambda} \, F_t^{(P)}
  & \ \approx \ -\frac{i \alpha_s C_F}{N_C} \, \frac{2 \pi f_\pi}{M_B} \, \frac{2}{\sqrt 3}
  \left( 1 + \frac{3  a_2^\pi}{2} 
  \right) \xi_\pi(\frac{M_B}{2}) \,,
  \label{Pwaverelation}
\end{align}
and 
\begin{align}
   F_0^{(D)} & \simeq \sqrt{\frac{2}{3}} \, F_\parallel^{(D)}
  \approx \frac{2 M_B \sqrt{q^2}}{\sqrt\lambda} \, F_t^{(D)}
  \cr 
  & \ \approx \ -\frac{i \alpha_s C_F}{N_C} \, 
  \frac{2 \pi f_\pi}{M_B} \, 
  \frac{\sqrt\lambda}{6\sqrt 5 M_B^2}
  \left( (5+6 a_2^\pi + \frac{2\mu_\pi}{M_B}) \, \xi_\pi(\frac{M_B}{2}) - (2+3 a_2^\pi) \, M_B \, \xi_\pi'(\frac{M_B}{2}) \right) \,,
  \cr &
  \label{Dwaverelation}
\end{align}
together with
\begin{align}
 F_\perp^{(P)} & \ \approx \
 \frac{i \alpha_s C_F}{N_C} \, \frac{2 \pi f_\pi}{M_B} \, \frac{\sqrt{3} \, \sqrt{\lambda}}{\sqrt 2 M_B^2}
  \left( 1 + a_2^\pi - \frac{\mu_\pi}{M_B} 
  \right) \xi_\pi(\frac{M_B}{2}) \,.
\end{align}

Notice that some of the above relations are a simple consequence of 
Lorentz invariance, as discussed in \cite{Hiller:2013cza}, 
since the number of independent 4-momentum vectors 
is reduced at the kinematic endpoint, $\sqrt\lambda \to 0$.
In particular, we recover in that limit
\begin{align}
  F_0  \simeq \cos\theta_\pi \, F_\parallel \left( 1 + {\cal O}\left(\frac{\sqrt\lambda}{M_B^2}\right)\right) 
  \label{F0rel} \,,
\end{align}
which implies  $F_\parallel^{(P)} \simeq \sqrt 2 \, F_0^{(P)}$,
$F_\parallel^{(D)} \simeq \frac{\sqrt 3}{\sqrt 2} \, F_0^{(D)}$ etc.

In order to assess the accuracy of the above relations, we study 
the form-factor ratios (properly normalized at $q^2\equiv 0$) 
as a function of the leptonic momentum transfer $q^2$.
The relations between the partial-wave projections
for the form factors $F_0$ and $F_t$ receive corrections of order 
$\sqrt{q^2}/M_B$ such that for $q^2 \sim 0.3$~GeV$^2$, the deviations 
from (\ref{Swaverelation} -- \ref{Dwaverelation}) are expected to be of
the order 10\%. This is indeed the case for the $S$- and $D$-wave,
while the corrections for the $P$-wave relation happen to imply large
numerical pre-factors which can be traced back to the slope of the 
$B \to \pi$ form factor at maximal recoil, $\xi_\pi'(M_B/2)$.
On the other hand, the relations between the  
partial-wave projections for $F_0$ and $F_\parallel$ are 
protected by Lorentz symmetry (\ref{F0rel}), and only receive small 
corrections of order $\sqrt{\lambda}/M_B^2$ which (in the kinematic
situation we are considering) scales as $q^2/M_B^2$. These relations
thus may still provide a reasonable approximation up to momentum transfers
of order $1$~GeV$^2$.

\subsection{Numerical results}

\newcommand{\MeV}{\text{MeV}}
\newcommand{\GeV}{\text{GeV}}
\begin{table}
\centering
\renewcommand{\arraystretch}{1.2}
\resizebox{\textwidth}{!}{
\begin{tabular}{|c| c| c| c| c|}
    \hline
    parameter                                   & value/interval
      & unit     & prior               & source/comments\\
    \hline
    \multicolumn{5}{|c|}{QCD input parameter}\\
    \hline
    $\alpha_s(m_Z)$                             &  0.1184  $\pm$ 0.0007
     & ---      & gaussian $@$ $68\%$ &  \cite{Beringer:1900zz}
   \\
    $\mu$                             &  $M_B/2$ $\pm$ $M_B/4$
      & \GeV     & gaussian$^\dagger$ $@$ $68\%$ &  \\
    $\overline{m}_{u+d}(2\,\GeV)$               &  7.8 $\pm$ 0.9
   & \MeV     & uniform $@$ $100\%$ & see \cite{Imsong:2014oqa}\\
    \hline
    \multicolumn{5}{|c|}{hadron masses}\\
    \hline
    $m_B$                                       &  5279.58
      & \MeV     & ---                 &  \cite{Beringer:1900zz}
     \\
    $m_\pi$                                     & 139.57
      & \MeV     & ---                 &  \cite{Beringer:1900zz}
     \\
    \hline
    \multicolumn{5}{|c|}{parameters of the pion DAs}\\
    \hline
    $f_\pi$                                     & $130.4$
     & \MeV     & ---                 &  \cite{Beringer:1900zz}
   \\
   $a_{2}^{\pi}(1\,\GeV)$                          & $[0.09, 0.25]$
      & ---      & uniform $@$ $100\%$ &  \cite{Khodjamirian:2011ub}
        \\
    $\mu_\pi(2\,\GeV)$                           & 2.5 $\pm$ 0.3
     & \GeV     & ---                 &  $m_\pi^2/(\overline{m}_{u+d})$ \\
    \hline
\end{tabular}
}
\caption{The input parameters that were used in our numerical analysis.
We express the prior distribution as a product of
individual priors
that are either uniform or gaussian. The uniform priors cover the stated
intervals with
100\% probability. The gaussian priors cover the stated intervals with 68\%
probability,
and the central value corresponds to the mode of the prior. For practical
purposes,
variates from the gaussian priors are only drawn from their respective 99\%
probability intervals. The prior for the parameters describing the $B\to \pi$
form factor $f_+$ are not listed here, and taken from \cite{Imsong:2014oqa}.
$\dagger$: We artificially restrict the support of the renormalization scale
$\mu$ to the interval $[M_B/4, M_B]$.
}
\label{tab:inputs}
\end{table}

In the following we will discuss numerical results for
\begin{itemize}
    \item the partial-wave expansion of the form factors,
    \item and two observables in the differential decay width of $B^-\to \pi^+\pi^-\mu^-\bar\nu_\mu$.
\end{itemize}
As already mentioned above, the corrections from spectator-scattering
encoded in $T_\Gamma^{\rm II}$ are a sub-leading effect and will be neglected for
simplicity.
Our prediction for the absolute values of the form factors and decay width is
still rather uncertain because of the overall factors of $\alpha_s(\mu)$ and
$\xi_\pi(E_2,\mu)$.
As we will see, a reduction of the uncertainties induced by $\xi_\pi$ and $\alpha_s$
can be achieved through suitable arithmetic combinations of form factors or observables.
For all numerical evaluations, we use the central values and uncertainty
intervals for the input parameters as listed in Table~\ref{tab:inputs},
as well as the correlated results of \cite{Imsong:2014oqa} for the
parameters describing the $B\to \pi$ form factor $f_+(\tilde{q}^2)$ in the region
$0 \leq \tilde{q}^2 \leq 12\,\GeV^2$. We find that the
uncertainties due to the soft-form-factor parameters are in all cases
smaller in size than the remaining parametric uncertainties, ranging from roughly
$30\%$--$90\%$ of the non-form-factor uncertainties.
(Note that we do not account for correlations
between the $B\to \pi$ form factor parameters and the parameters listed in
Table~\ref{tab:inputs}.)
The computations are made using the EOS software \cite{EOS}, which has been
extended for this purpose.

\begin{figure}[!h]
\centering
\begin{tabular}{cc}
    \includegraphics[width=.49\textwidth]{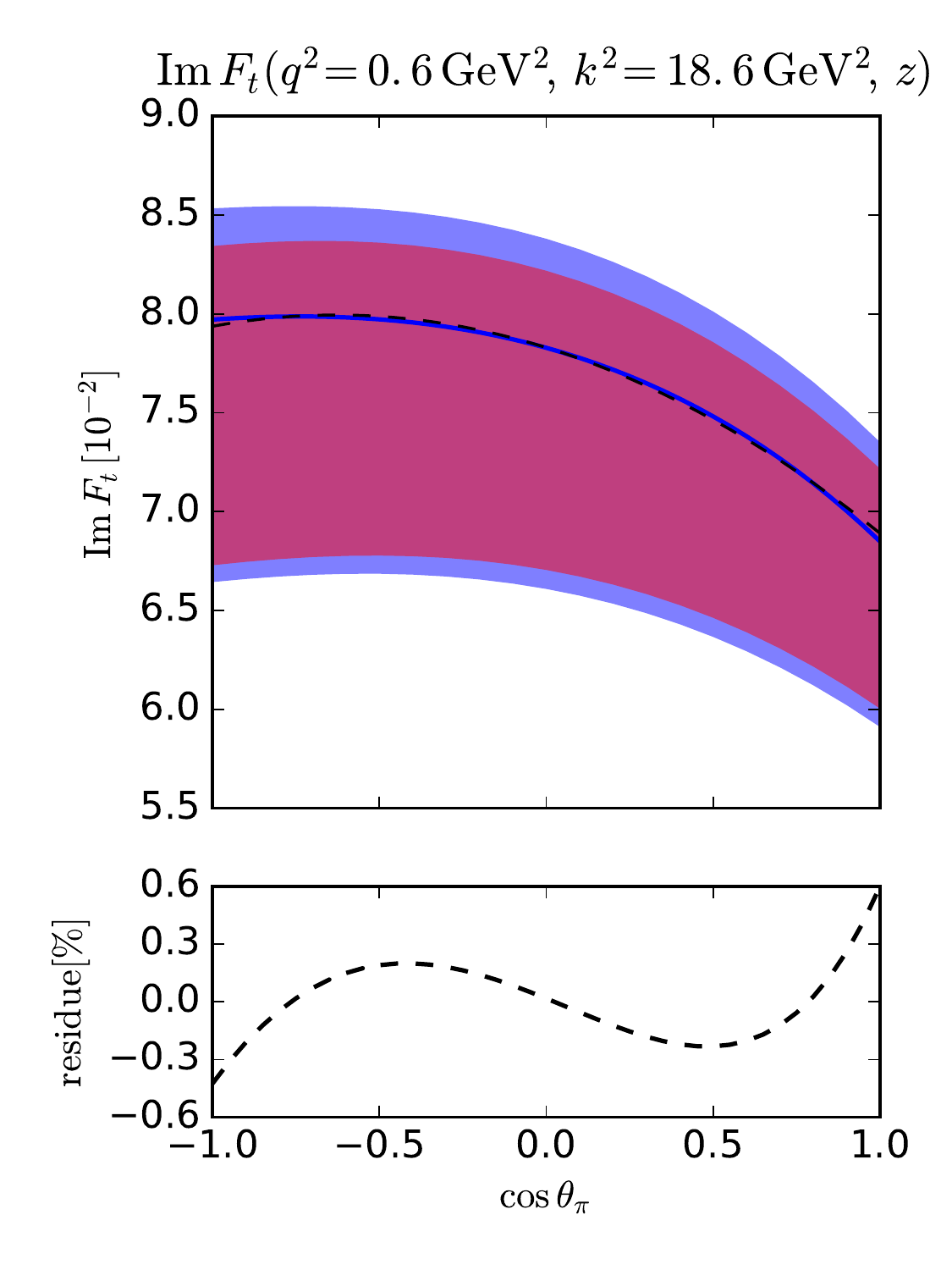} &
    \includegraphics[width=.49\textwidth]{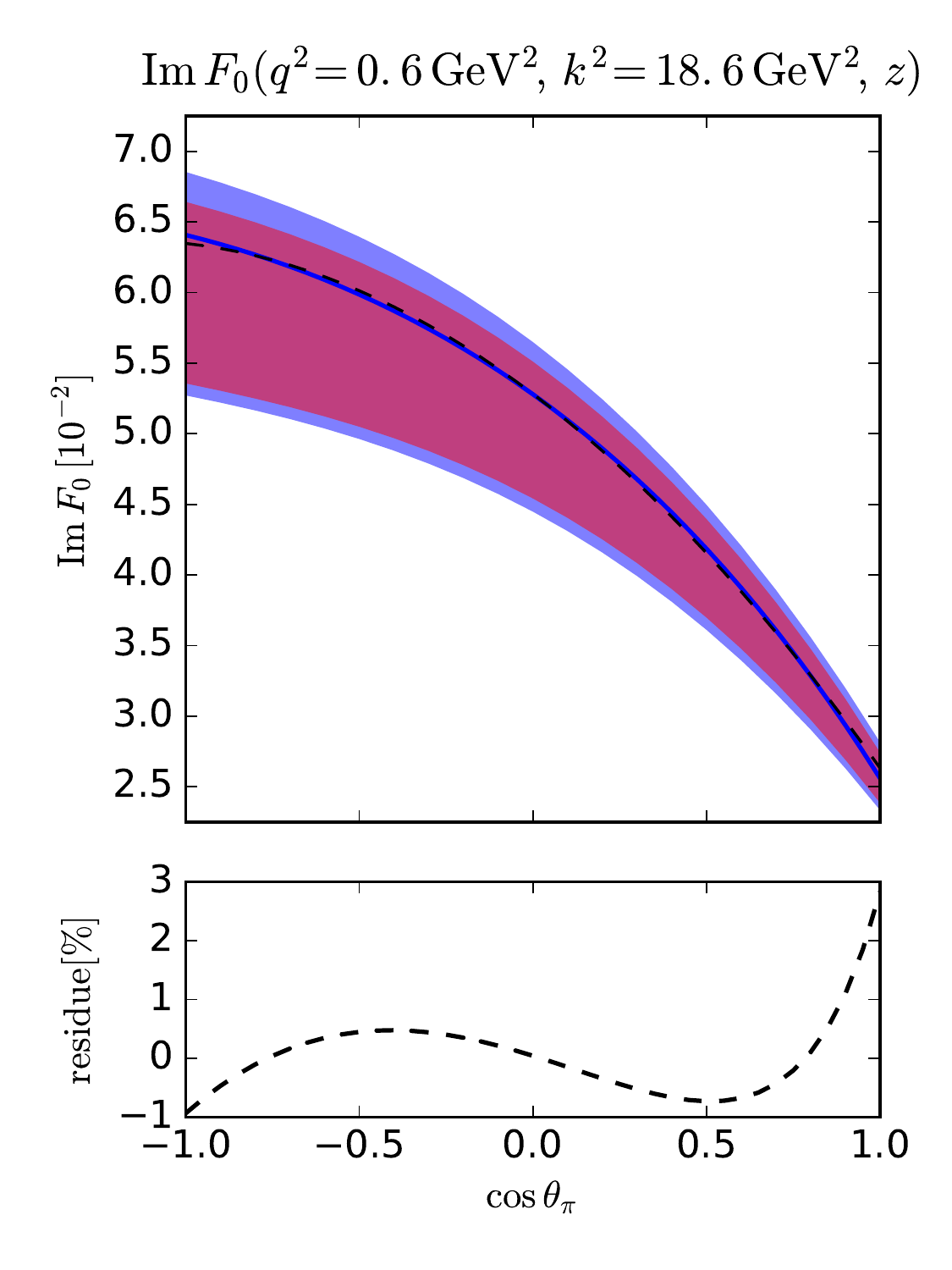} \\
    \includegraphics[width=.49\textwidth]{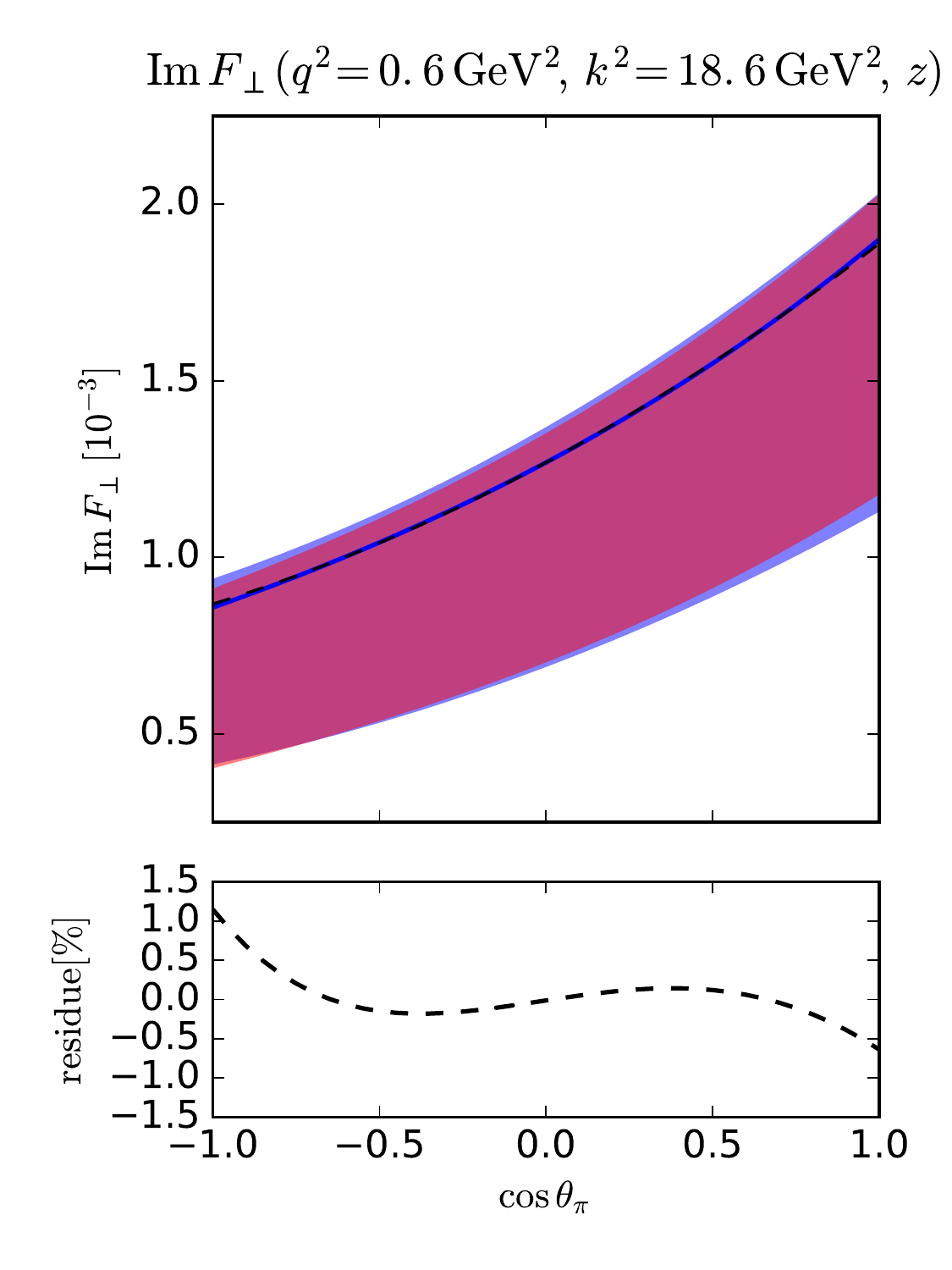} &
    \includegraphics[width=.49\textwidth]{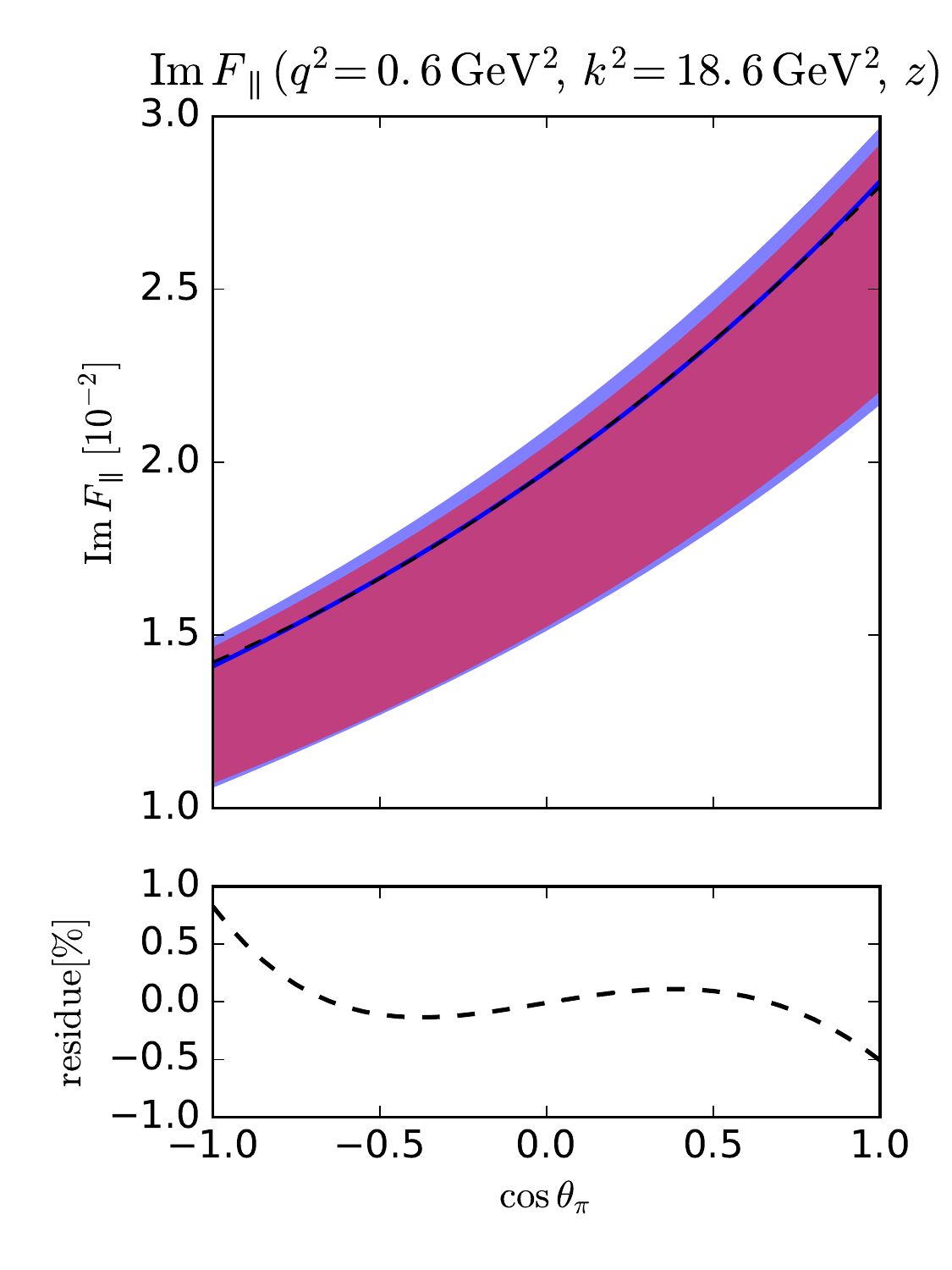}
\end{tabular}
\caption{
    Plots of the $\cos\theta_\pi$ dependence of the form factors
    in the phase space point $(q^2 = 0.6\,\text{GeV}^2, k^2 =
18.6\,\text{GeV}^2)$.
    The blue solid lines show the results at LO in $\alpha_s$, including both
    the twist-2 and twist-3 contributions. The blue shaded areas correspond to central
    $68\%$ intervals of the posterior-predictive distributions, which arise from the
    variation of the input parameters as listed in Table~\ref{tab:inputs}
    as well as the parameters for the $B\to \pi$ form factor $f_+$. The red shaded
    area is the same as the blue area, except for the $f_+$ variation.
    The black dashed lines show the
    approximation of each form factor by its first three partial waves. In the
    lower parts of each plot, the black dashed lines show the relative residue
    between the form factors and their partial-wave approximations.
    (Notice that in our convention, the form factors are purely imaginary at 
    leading order.)
}
\label{fig:partial-waves}
\end{figure}

\paragraph{Partial-wave expansion.}

We choose a benchmark point $(q^2 = 0.6\,\text{GeV}^2, k^2 =
18.6\,\text{GeV}^2)$, which corresponds to
\begin{align*}
    \frac{q^2}{M_B^2} & \approx 0.02, & \frac{\sqrt{\lambda}}{M_B^2} & \approx 0.20\,,
\end{align*}
in order to illustrate our results for the partial-wave expanded form
factors.
Each form factor is expanded up to its three leading partial waves, i.e.\
as a function of $z \equiv \cos\theta_\pi$, we have
\begin{align}
    F_{0(t)}^{S+P+D}(z)
        & = F_{0(t)}^S + \sqrt{3}\,F_{0(t)}^P z +
\sqrt{5}\,F_{0(t)}^D\,\frac{3 z^2 - 1}{2}\,,\\
    F_{\perp(\para)}^{P+D+F}(z)
    & = \sqrt{\frac{3}{2}}\,F_{\perp(\para)}^P +
\sqrt{\frac{15}{2}}\,F_{\perp(\para)}^D\,z +
\sqrt{\frac{21}{4}}\,F_{\perp(\para)}^F\,\frac{5 z^2 - 1}{2}\,,
\end{align}
where we have suppressed the $q^2$ and $k^2$ dependence of the form factors and
partial-wave coefficients for brevity.  One can now define relative residues
\begin{equation}
\begin{aligned}
    r_\lambda(z) & \equiv \frac{F_{\lambda}(z) -
F_{\lambda}^{S+P+D}(z)}{F_{\lambda}(z)}\,,\qquad \text{with\,}\lambda =
0,t\,,\\
    r_\lambda(z) & \equiv \frac{F_{\lambda}(z) -
F_{\lambda}^{P+D+F}(z)}{F_{\lambda}(z)}\,,\qquad \text{with\,}\lambda =
\perp,\para\,,
\end{aligned}
\end{equation}
in order to determine whether or not the form factors can be well approximated by
their partial wave expansion.  We find that
\begin{align}
    |r_0(z)|     & \leq 0.6\%\,,     & |r_t(z)|     & \leq 3.0\%\,,\\
    |r_\perp(z)| & \leq 1.2\%\,,     & |r_\para(z)| & \leq 0.8\%\,.
\end{align}
We therefore conclude that the first three partial waves approximate
the total $\cos\theta_\pi$ dependence of the form factors well.
These results are visualized in Fig.~\ref{fig:partial-waves}.

\paragraph{Decay width and pionic forward-backward asymmetry.}

Writing the 3-fold differential decay rate in terms of the kinematic
variables 
$(k^2,q^2, \cos\theta_\pi= \frac{2 \, q\cdot \bar k}{\sqrt\lambda})$,
we obtain in the SM
(for unexpanded 2-pion form factors, $F_i=F_i(k^2,q^2, q\cdot \bar k)$)\footnote{%
    Our result slightly disagrees with the $\beta_\ell$ dependence in Eqs.~(4.11) and (4.12) of \cite{Meissner:2013pba}
    in the arXiv version v2.
}
\begin{multline}
  \frac{d^3\Gamma(k^2, q^2, \cos\theta_\pi)}{dq^2 \, dk^2 \,
    d\cos\theta_\pi} \\
    = \frac{1}{4}
  |{\cal N}|^2 \beta_\ell
    \left[ (3 - \beta_\ell)|F_0|^2 + (1 -
    \cos^2\theta_\pi)(3 - \beta_\ell)\left(|F_\para|^2 +
|F_\perp|^2\right)
    + \frac{3 m_\ell^2}{q^2} |F_t|^2
    \right] \,,
\end{multline}
where the normalization factor reads
\begin{align}
    |{\cal N}|^2 & = G_F^2 \, |V_{ub}|^2 \, \frac{\beta_\ell \, q^2
\, \sqrt\lambda}{3 \cdot
    2^{10} \, \pi^{5} M_B^3} \,, &
    \text{with }& &
    \beta_\ell & = 1 - \frac{m_\ell^2}{q^2}\,.
\end{align}
\begin{table}[t]
    \renewcommand{\arraystretch}{1.2}
    \centering
    \begin{tabular}{|c|c|c|c r|}
        \hline
        & \multicolumn{4}{|c|}{result}\\
        phase space region & central & $\delta_\text{param}$ & $\delta_{f_+}$ & unit\\
        \hline
        \hline
        \multicolumn{5}{|c|}{$\mathcal{B}(B^-\to \pi^+\pi^-\mu^-\bar\nu_\mu)\,/\,|V_{ub}|^{2}$}\\
        \hline
        (A)   & $ 2.93$ & $^{+0.87}_{-0.40}$ & $^{+0.49}_{-0.35}$ & $10^{-8}$\\
        (B)   & $ 9.31$ & $^{+2.70}_{-1.30}$ & $^{+1.77}_{-0.69}$ & $10^{-7}$\\
        (A+B) & $ 9.60$ & $^{+2.80}_{-1.30}$ & $^{+1.89}_{-0.79}$ & $10^{-7}$\\
        (C)   & $ 3.18$ & $^{+0.63}_{-0.63}$ & $^{+0.48}_{-0.33}$ & $10^{-5}$\\
        \hline
        \hline
        \multicolumn{5}{|c|}{$A_\text{FB}^\pi(B^-\to \pi^+\pi^-\mu^-\bar\nu_\mu)$}\\
        \hline
        (A)   & $-1.96$ & $^{+0.15}_{-0.19}$ & $^{+0.04}_{-0.07}$ & $10^{-1}$\\
        (B)   & $-0.29$ & $^{+0.21}_{-0.19}$ & $^{+0.06}_{-0.11}$ & $10^{-1}$\\
        (A+B) & $-0.32$ & $^{+0.19}_{-0.21}$ & $^{+0.07}_{-0.11}$ & $10^{-1}$\\
        (C)   & $+1.25$ & $^{+0.07}_{-0.07}$ & $^{+0.03}_{-0.08}$ & $10^{-1}$\\
        \hline
    \end{tabular}
    \renewcommand{\arraystretch}{1.0}
    \caption{
        Numerical estimates for the partially-integrated branching ratio (in units of $|V_{ub}|^2$)
        and the pionic forward-backward asymmetry in different phase-space bins (see the text for more information).
        Note that our estimate for $A_\text{FB}^\pi$ in the region (C) has been obtained
        for $|\cos\theta_\pi| < 0.33$. The variation of all parameters, except the $B\to \pi$
        form factor $f_+$, comprise the uncertainty denoted as $\delta_\text{param}$. 
        The total uncertainty $\delta_\text{tot}$ is then obtained as
        $\delta_\text{tot}^2 = \delta_\text{param}^2 + \delta_{f_+}^2$.
        \label{tab:int-obs}
    }
\end{table}
\begin{figure}[t]
    \centering
    \includegraphics[width=.75\textwidth]{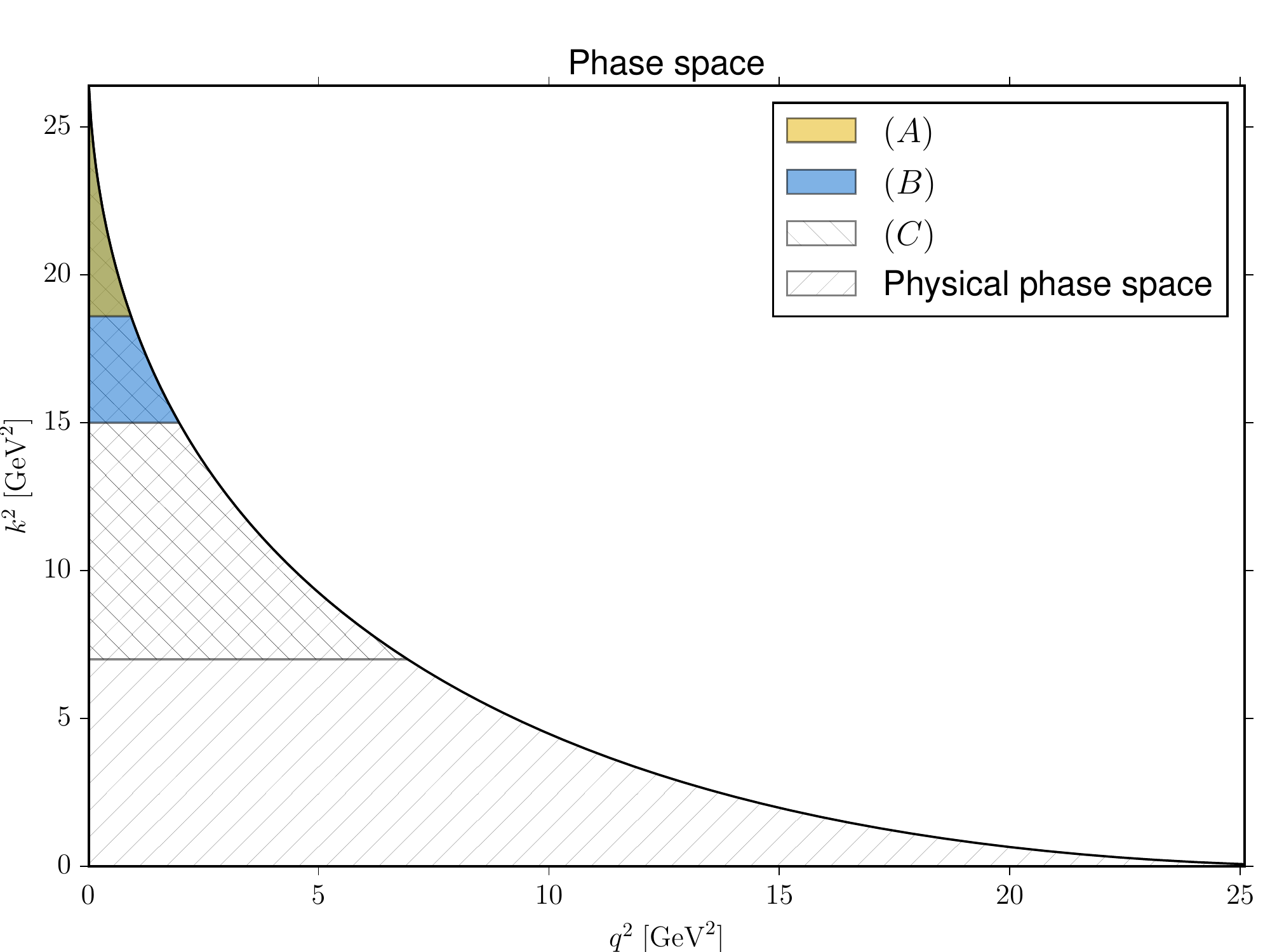}
    \caption{%
        We show our choices of phase space bins for the QCDF region
        (A: gold) and the extrapolation (B: blue). The region C,
        which has additionally limitations on the magnitude of $\cos\theta_\pi$, is illustrated
        as the `\textbackslash{}\textbackslash{}'-hatched region. The remainder of the physical phase space is
        highlighted as the `//'-hatched area. Estimates for the integrated
        $B^-\to \pi^+\pi^-\mu^-\bar\nu_\mu$ observables in different bins
        are shown in Table~\ref{tab:int-obs}.
        \label{fig:phase-space}
    }
\end{figure}
\begin{figure}[t]
    \centering
    \includegraphics[width=.75\textwidth]{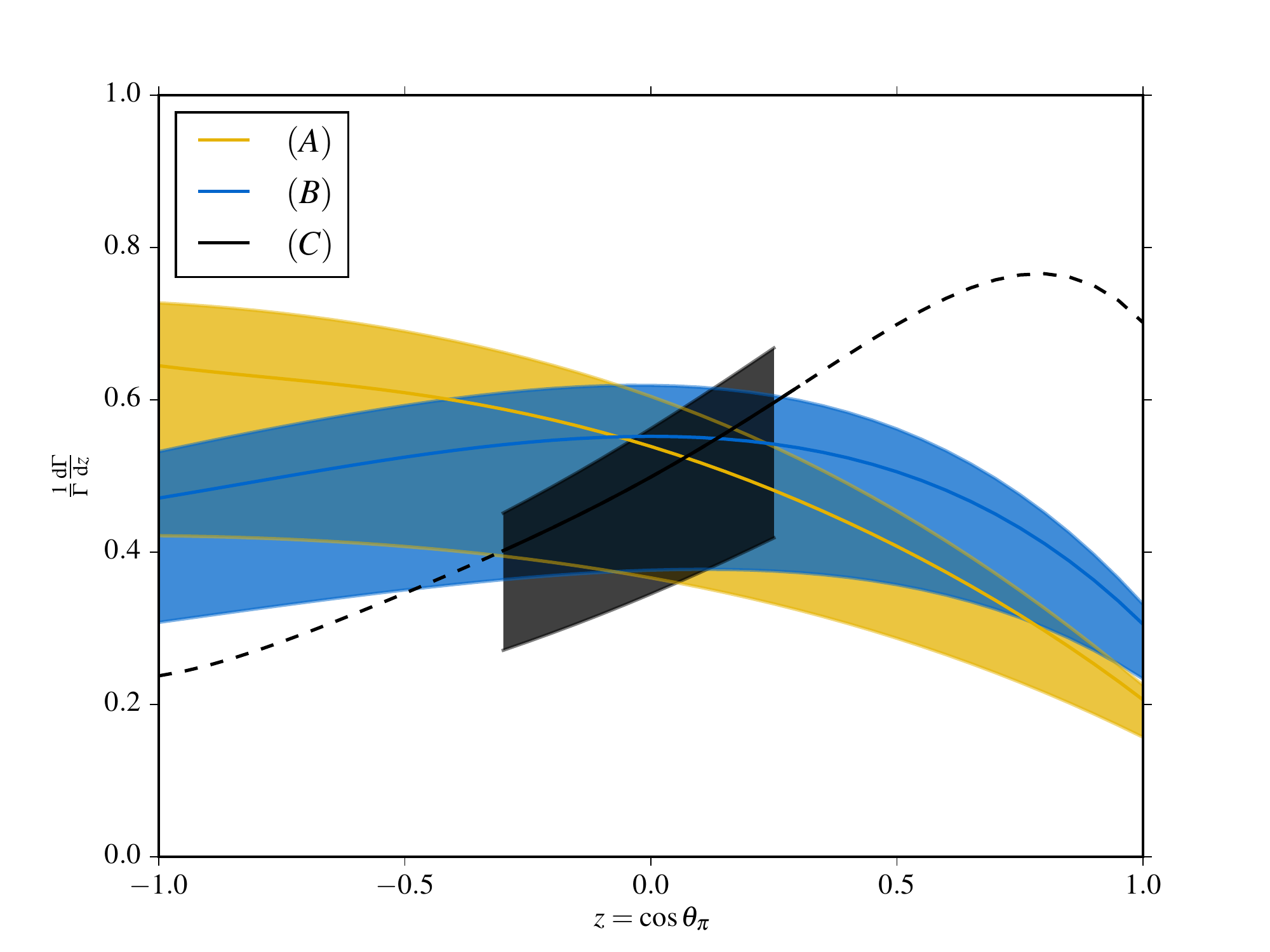}
    \caption{%
        Plot of the single-differential normalized decay rate as a function
        of $z \equiv \cos\theta_\pi$.
        The gold and blue shaded areas correspond to the phase space
        bins (A) and (B) as defined in the text. The bin (C)
        has additional restrictions on the size of $|z|$. An extrapolation
        beyond these restrictions is indicated by the dashed curve.
        The shaded areas correspond to the $68\%$ intervals as obtained
        from variation of all input parameters. The uncertainty is dominated by
        the parameters listed in Table~\ref{tab:inputs}.
        \label{fig:gamma-z}
    }
\end{figure}
The triple-differential branching ratio $\mathcal{B}(k^2,q^2, \cos\theta_\pi)$
can be used to define the two observables that we wish to discuss:
The partially-integrated branching ratio, as well as the pionic forward-backward
asymmetry for the decay:
\begin{equation}
    A^\pi_\text{FB}(k^2,q^2) \equiv \frac{\int_{-1}^{+1}
d\cos\theta_\pi\,\operatorname{sign}(\cos\theta_\pi)\,\mathcal{B}(k^2,q^2,
\cos\theta_\pi)}
    {\int_{-1}^{+1} d\cos\theta_\pi\,\mathcal{B}(k^2,q^2,
\cos\theta_\pi)}\,.
\end{equation}
In order to avoid controversies with the choice of the input value for
$|V_{ub}|$, we provide estimates for the branching ratio only in units of
$|V_{ub}|^{2}$. Due to the smallness of the differential branching ratio, we
prefer to provide our numerical estimates in form of binned observables. We
consider the three phase-space bins following from our discussion
in Sec.~\ref{sec:pc} for our numerical calculation (see also
Fig.~\ref{fig:phase-space} for a visualization in the $q^2$--$k^2$ plane):
\begin{align}
    \text{(A)} & :
        \begin{cases}
                0.02\,\GeV^2 \leq q^2 \leq (M_B - \sqrt{k^2})^2\,,\\
                18.60\,\GeV^2 \leq k^2 \leq (M_B - \sqrt{q^2})^2\,,\\
                -1 \leq \cos\theta_\pi \leq +1
        \end{cases}\\
    \text{(B)} & :
        \begin{cases}
                0.02\,\GeV^2 \leq q^2 \leq (M_B - \sqrt{k^2})^2\,,\\
                13.90\,\GeV^2 \leq k^2 \leq 18.60\,\GeV^2\,,\\
                -1 \leq \cos\theta_\pi \leq +1
        \end{cases}\\
    \text{(C)} & :
        \begin{cases}
                0.02\,\GeV^2 \leq q^2 \leq (M_B - \sqrt{k^2})^2\,,\\
                7.00\,\GeV^2 \leq k^2 \leq (M_B - \sqrt{q^2})^2\,,\\
                -0.33 \leq \cos\theta_\pi \leq +0.33
        \end{cases}
\end{align}
Region (A) corresponds to the phase space region in which the QCD-improved
factorization results are expected to hold rigorously. Region (B) extrapolates
to somewhat smaller values of $k^2$ (and the quoted uncertainties for this region
might be underestimated).
Finally, region (C) limits the phase space for the helicity angle
of the pions to $|\cos\theta_\pi| \leq 0.33$. This allows for using a larger
part of the $q^2$--$k^2$ plane, while still enforcing large pion energies in the
$B$ rest frame, $E_{1,2} > 1.24\,\GeV$.
Our results for both observables
are listed in Table~\ref{tab:int-obs}.
Moreover, we show the behaviour of the normalized single-differential
decay rate as a function of $\cos\theta_\pi$ in Fig.~\ref{fig:gamma-z}.
As can be seen, the decay features a sizeable pionic forward-backward
asymmetry in the phase-space bins (A) and (C).
Note, that the asymmetry switches sign when enlarging the phase space
toward bin (C). As a consequence, in the intermediate bin (B) the
asymmetry is one order of magnitude smaller than in either (A) or (C).

\clearpage 

\section{Summary}
\label{sec:summ}

In this work we have investigated the decay $B^- \to \pi^+\pi^-\ell^-\bar\nu_\ell$ 
in the context of QCD factorization (QCDF). To this end we have established a factorization
formula for $B\to\pi\pi$ form factors that is valid in the kinematic situation 
where both pions have large energy in the $B$-meson's rest frame with a large 
invariant dipion mass. The factorization formula takes a similar form as known
from other applications of the QCDF approach, with one term 
depending on a universal ``soft'' $B \to \pi$ form factor, and a second term
which completely factorizes in terms of hadronic light-cone distribution amplitudes (LCDAs). 
The leading contributions to the 
corresponding short-distance kernels $T^{\rm I}$ and $T^{\rm II}$ have been 
calculated for arbitrary Dirac structures of the underlying $b \to u$
transition current. 

To first approximation, 
all dipion form factors are proportional to the strong coupling $\alpha_s$ and
the soft $B\to \pi$ form factor, multiplied by linear combinations of 
only two independent  convolution integrals
involving the leading-twist LCDA of the positively
charged pion. This results in approximate relations between the 
dipion form factors and their partial-wave components which have been worked
out in detail. One class of corrections to the leading-order results arise from 
``chirally enhanced'' power corrections to $T^{\rm I}$.
Neglecting 3-particle Fock states in
the pion, the relevant twist-3 distribution amplitudes of the positively charged pion
are completely fixed, and therefore no additional hadronic unknowns arise.
The computation of the perturbative corrections due to spectator scattering,
which is described by the short-distance kernel $T^{\rm II}$, turns out to be more
involved. The final result appears as a consequence of a delicate cancellation 
of endpoint-divergent terms between the individual diagrams and the corresponding 
terms in the soft $B \to \pi$ form factor, providing a non-trivial confirmation
of the factorization formula to the considered order in the perturbative expansion.
The leading expression for $T^{\rm II}$  comes along with 
the first inverse moment of the $B$-meson LCDA.
On the light meson's side, we find somewhat more complicated convolution 
integrals. In the large-$N_C$ limit they reduce to three independent functions
that depend on the leading-twist pion LCDA.

In conclusion, the QCD factorization formula for $B \to \pi\pi$ form factors 
at large dipion mass and its implications are interesting from both, the theoretical
and phenomenological point of view. The factorization formula that we have 
established in this work combines features from semileptonic $B\to\pi$ transitions 
and non-leptonic $B \to \pi\pi$ decays with a non-trivial realization of the 
colour-transparency mechanism. Our results can also easily be generalized to other
decay modes like $B^- \to K^+ K^- \ell^-\bar \nu_\ell$ 
or $\bar B_s \to \pi^+ K^0 \ell^- \nu_\ell$.
Although the decay rate in the relevant 
kinematic region turns out to be too small to be of direct use for the 
determination of hadronic parameters or searches for new-physics effects, 
the approximate relations between the partial-wave form factors are useful
for the phenomenological modelling of $B \to \pi\pi\ell\nu$ decays over
the whole physical phase space.

\clearpage 

\section*{Acknowledgements}

We thank Bastian M\"uller for careful reading and checking of the manuscript.
This work is
supported in parts by the Bundesministerium f\"ur Bildung und Forschung (BMBF), and
by the Deutsche Forschungsgemeinschaft (DFG) within Research Unit FOR 1873 (“Quark
Flavour Physics and Effective Field Theories”).
D.v.D. acknowledges support by the Swiss National Science Foundation under grant
PP00P2-144674.

\appendix 

\section{Definition of Dipion Form Factors}
\label{app:ff}
\def\para{\parallel}

We follow the conventions in \cite{Faller:2013dwa}, and 
define vector and axial-vector form factors for $b \to u$ currents in the SM
as 
\begin{align}
    \langle \pi^+(k_1) \pi^-(k_2)|\bar\psi_{u}\gamma^\mu \psi_b|B^-(p)\rangle
    &= i F_\perp \, \frac{1}{\sqrt{k^2}} \  q_\perp^\mu
    \,,
    \label{eq:decay:hme:param1}
\end{align}
and 
\begin{align}
    -\langle \pi^+(k_1) \pi^-(k_2)|\bar\psi_{u}\gamma^\mu \gamma_5 \psi_b|B^-(p)\rangle  
    & = F_t \, \frac{q^\mu}{\sqrt{q^2}} 
    + F_0 \, \frac{2\sqrt{q^2}}{\sqrt{\lambda}} \, k_{(0)}^\mu
    + F_\parallel \, \frac{1}{\sqrt{k^2}} \, \bar{k}_{\para}^\mu\,,
\label{eq:decay:hme:param2}
\end{align}
where
\begin{equation}
\begin{aligned}
    k_{(0)}^\mu             & = k^\mu - \frac{k \cdot q}{q^2} \, q^\mu\,, \\
    \bar{k}_{\para}^\mu   & = \bar{k}^\mu - \frac{4 (k\cdot q) (q \cdot \bar{k})}{\lambda} \, k^\mu
                             + \frac{4 k^2 (q\cdot \bar{k})}{\lambda} \, q^\mu\,, \\
    q_{\perp}^\mu   &= 2 \, \epsilon^{\mu\alpha\beta\gamma} \, \frac{q_\alpha \, k_\beta \, \bar k_\gamma}{\sqrt{\lambda}}\,.
\end{aligned}
\label{eq:ortho}
\end{equation}
Here our convention for the Levi-Cevit\'a tensor is related to the definition of the
Dirac matrix $\gamma_5$ via 
\begin{align}
  {\rm tr} \left[\gamma_5 \, \gamma^\mu\gamma^\nu\gamma^\rho\gamma^\sigma \right]
  &= - 4 i \, \epsilon^{\mu\nu\rho\sigma} \,.
\end{align}
In terms of the so-defined ``helicity form factors'', one obtains simple expressions for 
the differential decay width and the angular observables, and simple relations between form factors 
in HQET or SCET, which has also been emphasized for other decay modes
\cite{Bharucha:2010im,Bobeth:2010wg,Feldmann:2011xf,Das:2012kz,Hambrock:2012dg}.
To extract the individual form factors, the above relations can be simply inverted,
\begin{equation}
\begin{aligned}
    F_\perp(k^2, q^2, q \cdot \bar{k}) & =- \frac{i \sqrt{k^2}}{q_\perp^2}  \,
    \langle \pi^+ \pi^-| \bar \psi_u \, \slashed q_\perp  \psi_b| B^-\rangle\,,\\
    F_\para(k^2, q^2, q \cdot \bar{k}) & = -\frac{\sqrt{k^2}}{\bar k_{\para}^2} 
    \, \langle \pi^+ \pi^-| \bar\psi_{u} \, \bar{\slashed k}_{\para} \gamma_5 \, \psi_b| B^-\rangle\,,\\
    F_0(k^2, q^2, q \cdot \bar{k})     & = - \frac{\sqrt\lambda}{2 \sqrt{q^2} \, k_{(0)}^2 }
    \, \langle \pi^+\pi^-|\bar\psi_{u} \, \slashed k_{(0)} \gamma_5 \, \psi_b|B^-\rangle\,,\\
    F_t(k^2, q^2, q \cdot \bar{k})     & = - \frac{1}{\sqrt{q^2}} \, \langle \pi^+\pi^-|\bar\psi_{u} \, \slashed q \gamma_5 \, \psi_b|B^-\rangle\,,
\end{aligned}
\label{A5}
\end{equation}
where
\begin{align} 
q_\perp^2 = 
   \bar k_\parallel^2 &= - \frac{ k^2 \, (4 E_1 E_2 -k^2)}{(E_1+E_2)^2-k^2}\,, \qquad  
   k_{(0)}^2 = -\frac{M_B^2 \, ((E_1+E_2)^2-k^2)}{q^2}\,.
\end{align}

These form factors can be further expanded in terms of partial waves
(see e.g.\ \cite{Faller:2013dwa}),
using
\begin{equation}
\begin{aligned}
    F_{\perp,\para}^{(\ell)}
    & = -\int_{-1}^{+1} \dd z\, \frac{\sqrt{2\ell + 1}}{2} F_{\perp,\para}(z)\, p_\ell^1(z)\sqrt{1 - z^2}\,,\\
    F_{0, t}^{(\ell)}
    & = +\int_{-1}^{+1} \dd z\, \frac{\sqrt{2\ell + 1}}{2} F_{0, t}(z)\, p_\ell^0(z)\,,
\end{aligned}
\end{equation}
where $p_\ell^m(z)$ denote the \emph{symmetrised} associated Legendre polynomials,
\begin{equation}
    p_\ell^m(z) \equiv \sqrt{\frac{(\ell - m)!}{(\ell + m)!}} P_\ell^m(z)\,,
\end{equation}
which fulfill the orthogonality relations
\begin{equation}
    \int_{-1}^{+1} \dd z\, p_\ell^m(z) p_k^m(z) = \frac{2}{2\ell + 1} \delta_{\ell k}\,,
\end{equation}
and $z \equiv \cos\theta_\pi = \frac{2 \, q\cdot \bar k}{\sqrt{\lambda}}$.
Notice that in our convention the form factors turn out to be purely
imaginary at leading order.

\section{Detailed Calculation for Kernel $T^{\rm II}$}

\label{app:t2}

In the following we summarize the individual results for the 
spectator-scattering diagrams that contribute to the kernel $T^{\rm II}$
at LO. We find it convenient to split the expressions into two terms:
one representing the individual contributions to the 
subprocess $b \to d \pi^+ g \ell^-\bar\nu_\ell$,
and the other the hard-collinear interaction with the spectator quark 
which induces the $B^- \to \pi^-$ transition, such that generically we 
have
\begin{align}
  \langle \pi\pi| \bar \psi_u \Gamma \psi_b|B\rangle \Big|_\text{Diagram X}
  &= {\rm tr}\left[ A_X \, A^{\rm spec} \right] \,, 
  \label{AXdef}
\end{align}
with
\begin{align}
A^{\rm spec} &= - g_s \, T^B \, \frac{{\cal M}_B \gamma_\beta {\cal M}_{\pi^-}}{(\ell - k_{\bar q2})^2} \simeq  g_s \, T^B \, \frac{{\cal M}_B \gamma_\beta {\cal M}_{\pi^-}}{2 \, \bar v \, \omega \, E_2}
\,.
\end{align}
Here the trace runs over Dirac and colour indices, and the 
integration over the (light-cone) momenta of the quarks is 
understood implicitly. The factor $(-i)$ from the hard-collinear gluon propagator
(in Feynman gauge) and the minus sign from the trace over 
the closed fermion loop has been assigned to the 
spectator term. If we restrict ourselves to the leading-power
contributions in the $1/m_b$ expansion, we can neglect the 
external transverse momenta in the \emph{hard} sub-process.
However, as is known and understood from the analogous case 
of $B \to \pi \ell \nu$ transitions \cite{Beneke:2000wa,Beneke:2003xr,Beneke:2003pa}, 
the impact of transverse
momenta in the hard-collinear spectator scattering is more 
subtle, and as a consequence transverse momenta in the 
associated propagator numerators must not be neglected from the
very beginning.
The resulting contribution to the $B \to \pi\pi$ matrix element 
will be decomposed according to (\ref{gdecomp}).

\subsection{Recapitulation: the $B \to \pi$ form factor $f_+$}

In this paper, we will use a physical definition of the 
soft $B \to \pi$ form factor $\xi_\pi(E_2)$.
To this end, we will identify it with the physical form
factor $f_+((p-k_2)^2)$, where $(p-k_2)^2=M_B^2 - 2 M_B E_2$.
The leading-power 
spectator-scattering contributions to $f_+$ have been 
calculated in \cite{Beneke:2000wa} and amount to 
\begin{align}
& \xi_\pi^{\rm (HSA)}(E_2) \equiv f_+^{\rm (HSA)}(E_2)
\cr 
&= 
\frac{\alpha_s}{4\pi} \, \frac{\pi^2 f_B f_\pi M_B}{N_C E_2^2}
\int_0^1 dv \int_0^\infty d\omega 
\left( g_+^{\rm finite}(v,\omega,E_2) + g_+^{\rm endpoint}(v,\omega,E_2) \right)
\,,
\label{xihsa}
\end{align}
with
\begin{align}
  g_+^{\rm finite}(v,\omega,E_2)  &= C_F \, \frac{4 E_2 - M_B}{M_B} \, 
  \frac{\phi_\pi(v)}{\bar v} \,  \frac{\phi_B^+(\omega)}{\omega} \,,
  \label{gplus}
\end{align}
and 
\begin{align}
  g_+^{\rm endpoint}(v,\omega,E_2)  &= C_F \,
   \frac{(1+\bar v)\, \phi_\pi(v)}{\bar v^2} \, \frac{\phi_B^-(\omega)}{\omega}  +
  2 \mu_\pi \, \frac{\phi_P(v)}{\bar v}\, \frac{\phi_B^+(\omega)}{\omega^2}
  \cr & \qquad + 
  \frac{\mu_\pi}{2E_2} \left( \frac{\phi_P(v)-\phi_\sigma'(v)/6}{\bar v^2} \right) \frac{\phi_B^+(\omega)}{\omega}
   \,.
\end{align}
Here the scaling of the various moments (after some ad-hoc regularization,
$\bar v \gtrsim \frac{\Lambda}{M_B}$, $\omega \gtrsim \frac{\Lambda^2}{M_B}$)
is to be understood as \cite{Beneke:2000wa}
\begin{align}
\left\langle \frac{\phi_\pi(v)}{\bar v} \right\rangle & \sim {\cal O}(1) \,, 
\cr 
 \left\langle \frac{\phi_P(v)+\phi_\sigma'(v)/6}{\bar v^2} \right\rangle \sim 
 \left\langle \frac{\phi_P(v)}{\bar v} \right\rangle \sim \left\langle \frac{\phi_\pi(v)}{\bar v^2} \right\rangle & \sim {\cal O}\left(\ln \frac{\Lambda}{M_B}\right) \,,
\cr  
\left\langle \frac{\phi_P(v)-\phi_\sigma'(v)/6}{\bar v^2} \right\rangle\sim 
 \left\langle \frac{\phi_P(v)}{\bar v^2} \right\rangle &\sim {\cal O}\left( \frac{M_B}{\Lambda} \right) \,,
\end{align}
and 
\begin{align}
\left \langle \frac{\phi_B^+(\omega)}{\omega} \right \rangle &= {\cal O}\left(\frac{1}{\Lambda} \right)
\,, 
\quad 
\left \langle \frac{\phi_B^+(\omega)}{\omega^2} \right \rangle \sim
{\cal O}\left(\frac{1}{\Lambda^2}
\, \ln \frac{\Lambda}{M_B} \right) \,, \quad 
\left \langle \frac{\phi_B^-(\omega)}{\omega} \right \rangle = {\cal O}\left(\frac{1}{\Lambda}
\, \ln \frac{\Lambda}{M_B} \right) \,.
\end{align}
In the following we have to show that the structures in $g_+^{\rm endpoint}$ are indeed
universal, and also appear in exactly the same form in 
the spectator-scattering contributions to the $B\to \pi\pi$
form factors at large $k^2$, justifying the procedure employed around 
(\ref{me1}).

\subsection{Expressions for $b \to d \pi^+g \ell^-\bar\nu_\ell$ amplitudes}

In the following, we collect the amplitudes $A_X$ describing the $b \to d \pi^+ g\ell^-\bar\nu_\ell$ subprocess in (\ref{AXdef}) 
from the various diagrams,
together with the approximations to be made in the large-recoil limit.

\vspace{1em}

\subsubsection{Diagrams (A1-A6)}

\begin{align}
A_1 &= 
 4\pi \alpha_s C_F \, g_s T^B\,  
 \frac{\gamma^\alpha \, {\cal M}_{\pi^+}^{(2)} \, \gamma_\alpha 
 \, (\slashed k_1 + \slashed k_{q2}) \, \Gamma \,
 (\slashed p - \slashed k_{\bar q2} + m_b) \, \gamma^\beta }{(k_1+ k_{q2})^2 (k_{q2}+k_{\bar q_1})^2 ((p_b - k_{\bar q2})^2-m_b^2)} 
 \cr & \simeq
 - 4\pi \alpha_s C_F \, g_s T^B \,  
 \frac{{\cal M}_{\pi^+}^{(2)} \, \slashed k_{2} \, \Gamma \,
 (\slashed p + M_B - \bar v \slashed k_{2}) 
 \, \gamma^\beta }{\bar u \, v \, \bar v \, M_B \, E_2 \, (k^2)^2} \,, \\[0.25em]
A_2 &= 
 4\pi \alpha_s C_F\, g_s T^B \,  
 \frac{\gamma^\beta \, (\slashed k_2 - \slashed \ell) \, 
   \gamma^\alpha \, {\cal M}_{\pi^+}^{(2)} \, \gamma_\alpha 
 \, (\slashed k - \slashed \ell) \, \Gamma}{ (k_2 - \ell)^2(k - \ell)^2 ( k - \ell -k_{q_1})^2}
 \cr &
 \simeq
 -4\pi \alpha_s C_F \,g_s T^B \,
 \frac{\gamma^\beta \, (\slashed k_2 - \slashed \ell) \, {\cal M}_{\pi^+}^{(2)} \, \slashed k_{2} \, \Gamma}{\bar u \, \omega \, E_2\, (k^2)^2} \,,
 \end{align}
and 
\begin{align}
A_3 &= - 4 \pi \alpha_s  \, C_{FA}  \, g_s T^B \,
\frac{\gamma^\alpha {\cal M}_{\pi^+}^{(2)} \gamma^\beta 
(\slashed k_{q1} + \slashed k_{\bar q2} - \slashed \ell) \gamma_\alpha  (\slashed k - \slashed \ell ) \Gamma}{(k_{q1} + k_{\bar q2} - \ell)^2((k-\ell)^2)(k_{\bar q1} + k_{q2})^2}
\cr 
&\simeq 
- 4 \pi \alpha_s \, C_{FA} \, g_s T^B \,
\frac{\gamma^\alpha {\cal M}_{\pi^+}^{(2)} \gamma^\beta 
(u \slashed k_{1} + \bar v \slashed k_{2}) \gamma_\alpha \slashed k \, \Gamma}{
u \bar u \, v \bar v \, (k^2)^3} \,,
\\[0.25em]
A_4 &=  - 4 \pi \alpha_s \, C_{FA} \, g_s T^B \,
\frac{\gamma^\alpha (\slashed l - \slashed k_{\bar q1}-\slashed k_{\bar q2}) \, \gamma^\beta
{\cal M}_{\pi^+}^{(2)} \gamma_\alpha  (\slashed k -\slashed \ell) \Gamma}{(\ell- k_{\bar q1} - k_{\bar q2})^2((k-\ell)^2)(k_{\bar q1} + k_{2}-\ell)^2}
\cr 
&\simeq 
4 \pi \alpha_s \, C_{FA} \, g_s T^B \,
\frac{\gamma^\alpha (\bar u \slashed k_1 + \bar v \slashed k_2) \, \gamma^\beta
{\cal M}_{\pi^+}^{(2)}  \gamma_\alpha \slashed k \, \Gamma}{
\bar u^2 \, \bar v \, (k^2)^3} \,,
\end{align}
and 
\begin{align}
A_5 &=  4 \pi \alpha_s  \, C_F \, g_s T^B \,
\frac{\gamma^\alpha {\cal M}_{\pi^+}^{(2)} \gamma_\alpha 
(\slashed k_1 + \slashed k_{q2}) \gamma^\beta  (\slashed k -\slashed \ell) \Gamma}{(k_1+k_{q2})^2 ((k-\ell)^2)(k_{\bar q1} + k_{q2})^2}
\cr 
&\simeq 
8 \pi \alpha_s \, C_F \, g_s T^B \,
\frac{{\cal M}_{\pi^+}^{(2)} \slashed k_{2} \gamma^\beta \slashed k \, \Gamma}{
\bar u \, v \, (k^2)^3} \,,
\end{align}
and
\begin{align}
A_6 &=  4 \pi \alpha_s  \, \frac{C_A}{2} \, g_s T^B \, \frac{\gamma_\alpha {\cal M}_{\pi^+}^{(2)} \gamma_\gamma (\slashed k - \slashed \ell) \Gamma}{
(k-\ell)^2 (k_{\bar q1} + k_2 - \ell)^2 (k_{\bar q1}+k_{q2})^2}
\cr & \qquad \times \left( 
 g^{\alpha\beta} (k_{\bar q2}-k_{q2}-k_{\bar q1}- \ell)^\gamma 
 +
 g^{\beta\gamma} (2\ell - k_{\bar q1} - k_2 - k_{\bar q2})^\alpha 
  \right. 
 \cr 
 & \left. \qquad \qquad +
 g^{\alpha\gamma} (2 k_{\bar q1} + k_2 + k_{q2} - \ell)^\beta
\right)
\cr 
&\simeq 
2 \pi \alpha_s \, C_A \, g_s T^B \,
\frac{\gamma_\alpha {\cal M}_{\pi^+}^{(2)} \gamma_\gamma \slashed k \, \Gamma}{
\bar u^2 \, v \, (k^2)^3}
\cr & \qquad \times 
\left(
g^{\alpha\beta}  (\bar v-v) k_2^\gamma 
-
 g^{\beta\gamma} (1+\bar v)\, k_2^\alpha 
+
g^{\alpha\gamma} \, (2\bar u k_1 + (1+v) \, k_2)^\beta
\right) 
\,.
\end{align}

\vspace{1em}

\subsubsection{Diagrams (B1-B6)}

\begin{align}
B_1 &= 
4\pi \alpha_s C_F \, g_s T^B\, 
\frac{ \gamma^\alpha \, {\cal M}_{\pi^+}^{(2)} \, \Gamma \,
(\slashed p - \slashed k_{\bar q1} - \slashed k_2 + m_b) \, 
\gamma_\alpha \, 
(\slashed p - \slashed k_{\bar q2}+m_b) \, \gamma^\beta}
{((p-k_{\bar q1}- k_2)^2-m_b^2) \, ((p-k_{\bar q2})^2-m_b^2) \, (k_{\bar q1}+k_{q2})^2}
 \cr & \simeq
 - 4\pi \alpha_s C_F \, g_s T^B \,  
 \frac{\gamma^\alpha \, {\cal M}_{\pi^+}^{(2)} \, \Gamma \,
 (\slashed p  -\bar u\slashed k_1- \slashed k_2+M_B)\,
 \gamma_\alpha\, (\slashed p - \bar v\slashed k_2+ M_B) \, \gamma^\beta}{
 2 \bar u v \bar v \,  E_2 M_B  \, k^2 \,(- 2 \bar u E_1 M_B-2 E_2 M_B  + \bar u k^2)} \,,
 \cr &
 \\
B_2 &= 
4\pi \alpha_s C_F \, g_s T^B\, 
\frac{ \gamma^\beta \, (\slashed k_2-\slashed l) \,
\gamma^\alpha \, {\cal M}_{\pi^+}^{(2)} \, \Gamma \,
(\slashed p - \slashed k_{\bar q1} - \slashed k_2 -\slashed l+ m_b) \, 
\gamma_\alpha}
{(k_2-\ell)^2 \, ((p-k_{\bar q1}- k_2-\ell)^2-m_b^2) \, (k_{\bar q1}+k_{2}-\ell)^2}
 \cr & \simeq
 - 4\pi \alpha_s C_F \, g_s T^B \, 
\frac{ \gamma^\beta \, (\slashed k_2-\slashed l) \,
\gamma^\alpha \, {\cal M}_{\pi^+}^{(2)} \, \Gamma \,
(\slashed p - \bar u\slashed k_1 - \slashed k_2 +M_B) \, 
\gamma_\alpha}
{2 \bar u \, \omega  \, E_2 \,k^2 \, (-2 \bar u E_1 M_B - 2 E_2 M_B + \bar u k^2) } \,,
\end{align}
and
\begin{align}
B_3 &=  -4 \pi \alpha_s  \, C_{FA} \, g_s T^B \,
\frac{\gamma^\alpha {\cal M}_{\pi^+}^{(2)} \gamma^\beta 
(\slashed k_{q1} + \slashed k_{\bar q2} - \slashed \ell) \Gamma 
(\slashed p -\slashed k_{\bar q1}-\slashed k_{q2} - \slashed \ell + m_b) \gamma_\alpha}{
(k_{q1} + k_{\bar q2} - \ell)^2 
((p-k_{\bar q1} - k_{q2}-\ell)^2-m_b^2) (k_{\bar q1}+k_{q_2})^2}
\cr 
&\simeq 
- 4 \pi \alpha_s \, C_{FA} \, g_s T^B \,
\frac{\gamma^\alpha {\cal M}_{\pi^+}^{(2)} \gamma^\beta 
(u \slashed k_{1} + \bar v \slashed k_{2}) \Gamma 
(\slashed p - \bar u \slashed k_{1}- v \slashed k_{2}  + M_B) \gamma_\alpha}{
u\bar u \, v \bar v \, (k^2)^2 \, 
(-2 \bar u E_1  M_B - 2 v E_2 M_B + \bar u v k^2)}
\,,
\\
B_4 &=  - 4 \pi \alpha_s  \, C_{FA} \, g_s T^B \,
\frac{\gamma^\alpha 
(-\slashed k_{\bar q1} -\slashed k_{\bar q2} + \slashed \ell) \gamma^\beta
{\cal M}_{\pi^+}^{(2)} \Gamma 
(\slashed p -\slashed k_{\bar q1}-\slashed k_{2} + m_b) \gamma_\alpha}{
(k_{\bar q1} + k_{\bar q2} - \ell)^2 
((p-k_{\bar q1} - k_{2})^2-m_b^2) (k_{\bar q1}+k_{2}-\ell)^2}
\cr 
&\simeq 
- 4 \pi \alpha_s \, C_{FA} \, g_s T^B \,
\frac{\gamma^\alpha 
(-\bar u \slashed k_{1} -\bar v \slashed k_{2}) \gamma^\beta
{\cal M}_{\pi^+}^{(2)} \Gamma 
(\slashed p -\bar u \slashed k_{1}-\slashed k_{2} + M_B) \gamma_\alpha}{
\bar u^2 \, \bar v \, (k^2)^2 \,
(-2 \bar u E_1 M_B - 2 E_2 M_B + \bar u k^2)}
\,,
\cr &
\\
B_5 &= - 4 \pi \alpha_s  \, C_{FA}\, g_s T^B \,
\frac{\gamma^\alpha  {\cal M}_{\pi^+}^{(2)} \Gamma (\slashed p- \slashed k_{\bar q1}-\slashed k_2 + m_b) \gamma^\beta 
(\slashed p- \slashed k_{\bar q1}-\slashed k_{q2}-\slashed \ell  + m_b)\gamma_\alpha}{((p-k_{\bar q1}-k_2)^2-m_b^2)
((p-k_{\bar q1}-k_{q2}- \ell)^2-m_b^2)(k_{\bar q1}+k_{q2})^2}
\cr 
&\simeq  
- 4 \pi \alpha_s  \, C_{FA} \, g_s T^B \,
\cr & \qquad \times 
\frac{\gamma^\alpha  {\cal M}_{\pi^+}^{(2)} \Gamma (\slashed p- \bar u \slashed k_{1}-\slashed k_2 + M_B) \gamma^\beta 
(\slashed p- \bar u \slashed k_{1}-v \slashed k_{2} + M_B)\gamma_\alpha}{
(-2 \bar u M_B E_1 - 2 M_B E_2 + \bar u k^2)
(-2 \bar u M_B E_1 - 2 v M_B E_2 + \bar uv k^2) \, \bar u v k^2}
\,,
\end{align}
and 
\begin{align}
B_6 &=  4 \pi \alpha_s  \, \frac{C_A}{2} \, g_s T^B \, \frac{\gamma_\alpha {\cal M}_{\pi^+}^{(2)} 
\Gamma (\slashed p - \slashed k_{\bar q1} - \slashed k_2+m_b) 
\gamma_\gamma}{
((p-k_{\bar q1}-k_2)^2- m_b^2) (k_{\bar q1} + k_2 - \ell)^2 (k_{\bar q1}+k_{q2})^2}
\cr & \qquad \times \left( 
 g^{\alpha\beta} (k_{\bar q2}-k_{q2}-k_{\bar q1}- \ell)^\gamma 
 +
 g^{\beta\gamma} (2\ell - k_{\bar q1} - k_2 - k_{\bar q2})^\alpha 
  \right. 
 \cr 
 & \left. \qquad \qquad +
 g^{\alpha\gamma} (2 k_{\bar q1} + k_2 + k_{q2} - \ell)^\beta
\right)
\cr 
&\simeq 
2 \pi \alpha_s \, C_A \, g_s T^B \,\frac{\gamma_\alpha {\cal M}_{\pi^+}^{(2)} 
\Gamma (\slashed p - \bar u \slashed k_{1} - \slashed k_2+M_B) 
\gamma_\gamma}{
(- 2 \bar u E_1 M_B - 2 E_2 M_B + \bar u k^2) \, \bar u^2 v \, (k^2)^2}
\cr & \qquad \times \left( 
 g^{\alpha\beta} \, ((\bar v-v)\, k_{2}-\bar u k_{1})^\gamma 
 -
 g^{\beta\gamma} \, (1+\bar v) \, k_2^\alpha 
  +
 g^{\alpha\gamma} \, ( 2 \bar u k_{1} + v k_2)^\beta
\right)
\,.
\end{align}

\vspace{1em}

\subsection{Contributions to $B \to \pi\pi$ matrix elements}

In the following we collect the finite and endpoint divergent
contributions of the individual Feynman diagrams to the 
$B \to \pi\pi$ matrix elements as defined in Eq.~(\ref{gdecomp}).
The contributions to the kernel $T_\Gamma^{\rm II}$ from 
the spectator scattering diagrams are expressed in terms of
several functions of the momentum fractions $\bar u$ and $\bar v$
of the (anti-)quarks in the two pions which are convoluted 
with the corresponding leading-twist LCDAs. 
In the following we use the same abbreviations
for Dirac traces (\ref{sdef}), kinematic invariants (\ref{vperpdef}),
colour factors (\ref{CFAdef}) 
as defined in the main body of the article.
We also employ the equations of motion (\ref{phi3effdef}) to simplify the 
twist-3 contributions to the endpoint-divergent terms in 
the hard-scattering amplitudes.

\vspace{1em}

\subsubsection{Diagram (A1)}
\begin{align}
g_{(A1)}^{\rm finite} &= C_F \, 
\, \frac{E_2}{M_B} \, \frac{s_4}{\bar u} \, 
 \frac{\phi_\pi(v)}{v\bar v} \, \frac{\phi_B^+(\omega)}{\omega} 
\end{align}
and 
\begin{align}
g_{(A1)}^{\rm endpoint} &= C_F \, 
\frac{ S_A}{\bar u} 
\left( \frac{\phi_\pi(v)}{v\bar v^2}  \frac{\phi_B^-(\omega)}{\omega}
+  \frac{\mu_\pi \phi_\sigma(v)}{6 E_2 \,\bar v^3 } \,  \frac{\phi_B^+(\omega)}{\omega}
\right)
\end{align}

\vspace{1em}

\subsubsection{Diagram (A2)}
\begin{align}
g_{(A2)}^{\rm finite} &= C_F
\left( \frac{2 E_2 M_B}{k^2} \, \frac{s_6}{\bar u} - \frac{s_2}{\bar u}\right) 
  \frac{\phi_\pi(v)}{\bar v} \,
  \frac{\phi_B^+(\omega)}{\omega} \,,
\end{align}
and 
\begin{align}
g_{(A2)}^{\rm endpoint} &= C_F \, 
\frac{S_A}{\bar u} 
\left( \frac{ \phi_\pi(v)}{\bar v}\, \frac{\phi_B^-(\omega)}{\omega}
+ 2 \mu_\pi \, \frac{\phi_P(v)}{\bar v}\, \frac{\phi_B^+(\omega)}{\omega^2}
\right) \,.
\end{align}

\vspace{1em}

\subsubsection{Diagrams (A3+A4)}
\begin{align}
g_{(A3+A4)}^{\rm finite} &= C_{FA} \,
\left( \frac{2 E_2 M_B}{k^2} \, \frac{s_6}{\bar u}- \frac{s_2}{\bar u} 
 \right)  \frac{\phi_\pi(v)}{v\bar v} \, \frac{\phi_B^+(\omega)}{\omega}  \,,
\end{align}
and 
\begin{align}
g_{(A3+A4)}^{\rm endpoint} &= -C_{FA} \left( 
\frac{2 E_2 M_B}{k^2} \, \frac{s_5}{\bar u^2} 
\, \frac{ \phi_\pi(v)}{\bar v} \, \frac{\phi_B^+(\omega)}{\omega}
- \frac{S_A}{\bar u}
\, \frac{ \phi_\pi(v)}{v\bar v} \, \frac{\phi_B^-(\omega)}{\omega} \right) \,.
\end{align}

\vspace{1em}

\subsubsection{Diagram (A5)}
\begin{align}
g_{(A5)}^{\rm finite} &= C_F \, \frac{2 E_2 M_B}{k^2} \, \frac{s_5}{\bar u} 
 \, 
  \frac{\phi_\pi(v)}{v\bar v} \, \frac{\phi_B^+(\omega)}{\omega}  \,,
\end{align}
and 
\begin{align}
g_{(A5)}^{\rm endpoint} &= 0  \,.
\end{align} 

\vspace{1em}

\subsubsection{Diagram (A6)}
\begin{align}
g_{(A6)}^{\rm finite} &= C_A \left(\frac{s_2}{\bar u} - \frac{2 E_2 M_B}{k^2} 
\, \frac{s_6}{\bar u} 
 \right)
  \frac{\phi_\pi(v)}{2 v \bar v} \,\frac{\phi_B^+(\omega)}{\omega}  \,,
\end{align}
and 
\begin{align}
g_{(A6)}^{\rm endpoint} &= C_A \left( 
\frac{2 E_2 M_B}{k^2} \, \frac{s_5}{\bar u^2} 
\, \frac{ (v-\bar v) \, \phi_\pi(v)}{4 v \bar v} \, \frac{\phi_B^+(\omega)}{\omega} 
- \frac{S_A}{\bar u} 
 \, 
  \frac{\phi_\pi(v)}{2v\bar v}  \frac{\phi_B^-(\omega)}{\omega} \right) \,.
\end{align}

\vspace{1em}

\subsubsection{Diagram (B1)}
\begin{align}
g_{(B1)}^{\rm finite} &= C_F \, 
\, \frac{2 E_2}{M_B} \, 
\frac{S_B^{\rm (i)}(u)}{\bar u}
  \, \frac{\phi_\pi(v)}{v \bar v} \,\frac{\phi_B^+(\omega)}{\omega}  \,,
\end{align}
and 
\begin{align}
g_{(B1)}^{\rm endpoint} &= C_F \, 
\frac{ S_B^{\rm (i)}(u)+S_B^{\rm (ii)}(u)}{\bar u} 
\left( \frac{\phi_\pi(v)}{ v\bar v^2} \, \frac{\phi_B^-(\omega)}{\omega}
+  \frac{\mu_\pi \phi_\sigma(v)}{6 E_2 \, \bar v^3} \, \frac{\phi_B^+(\omega)}{\omega}
\right) \,.
\end{align}

\vspace{1em}

\subsubsection{Diagram (B2)}
\begin{align}
g_{(B2)}^{\rm finite} &= C_F \left(
\left(\frac{2 E_2}{M_B} - 1 \right) \frac{ S_B^{\rm (i)}(u)}{\bar u}
+ \frac{E_2}{M_B} \, \frac{s_3}{\bar u}
  \right) \frac{\phi_\pi(v)}{\bar v} \, \frac{\phi_B^+(\omega)}{\omega}  \,,
\end{align}
and 
\begin{align}
g_{(B2)}^{\rm endpoint} &= C_F \, 
\frac{ S_B^{\rm (i)}(u)+S_B^{\rm (ii)}(u)}{\bar u} 
\left( 
 \frac{ \phi_\pi(v)}{\bar v} \, \frac{\phi_B^-(\omega)}{\omega}
+  2 \mu_\pi \, \frac{\phi_P(v)}{\bar v} \, \frac{\phi_B^+(\omega)}{\omega^2}
\right)\,. 
\end{align}

\vspace{1em}

\subsubsection{Diagrams (B3+B5)}

\begin{align}
g_{(B3+B5)}^{\rm finite} &=
C_{FA} \left(
-v_\perp^2 \, \frac{S_B^{(i)}(u)}{\bar u} 
\left(1 - \frac{2 E_2 M_B}{\bar u v k^2 - 2 \bar u E_1 M_B - 2 v E_2 M_B} \right)
\right. 
\cr 
& \left. \phantom{C_{FA}} \qquad 
+ \frac{ 2 E_2^2 \, s_3 - E_2 \, M_B 
\, (2 s_5- s_7)}{\bar u \, (\bar u v k^2 - 2 \bar u E_1 M_B - 2 v E_2 M_B)}
\right) \frac{\phi_\pi(v)}{v\bar v} \, \frac{\phi_B^+(\omega)}{\omega}
\,,
\end{align}
and 
\begin{align}
g_{(B3+B5)}^{\rm endpoint} &= C_{FA} 
\left( 
\frac{S_B^{(i)}(u)+S_B^{(ii)}(u)}{\bar u} \, \frac{\phi_\pi(v)}{v\bar v^2} \, 
\frac{\phi_B^-(\omega)}{\omega}
-v_\perp^2 \, \frac{S_B^{(i)}(u)}{\bar u} 
 \frac{ \phi_\pi(v)}{\bar v^2} \, \frac{\phi_B^+(\omega)}{\omega}
\right. 
\cr & \qquad\phantom{ C_{FA}}
\left. 
+
v_\perp^2 \, \frac{S_B^{ (i)}(u)}{\bar u} 
 \frac{ \mu_\pi \phi_\sigma(v)}{6 E_2 \bar v^3} \, \frac{\phi_B^-(\omega)}{\omega}
\right. 
\cr & \qquad\phantom{ C_{FA}}
\left. 
- v_\perp^2 \, \frac{S_B^{(i)}(u)+S_B^{(ii)}(u)}{\bar u} 
 \frac{ \mu_\pi \phi_\sigma(v)}{6 E_2 \bar v^3} \, \frac{\phi_B^+(\omega)}{\omega}
\right) \,.
\end{align}

\vspace{1em}

\subsubsection{Diagram (B4)}
\begin{align}
g_{(B4)}^{\rm finite} &= 
 C_{FA} \left(
 \frac{2 E_2}{M_B} \left( \frac{2 E_1 M_B}{k^2}-1 \right) \frac{S_B^{(i)}(u) }{\bar u } 
 - \frac{E_2}{M_B} \, \frac{s_3}{\bar u} \right)
 \frac{\phi_\pi(v)}{\bar v} \, \frac{\phi_B^+(\omega)}{\omega}
\,,
\end{align}
and 
\begin{align}
g_{(B4)}^{\rm endpoint} &= C_{FA} 
\left( 
\frac{2 E_2 M_B}{k^2}\, \frac{s_5}{\bar u^2} 
\, \frac{\phi_\pi(v)}{\bar v} \, \frac{\phi_B^+(\omega)}{\omega} 
- \frac{S_B^{(i)}(u)+S_B^{(ii)}(u)}{\bar u} \, \frac{\phi_\pi(v)}{\bar v^2} \, 
\frac{\phi_B^-(\omega)}{\omega}
\right. 
\cr & \qquad\phantom{ C_{FA}}
\left. 
+ v_\perp^2 \, \frac{S_B^{(i)}(u)}{\bar u} 
 \frac{ \phi_\pi(v)}{\bar v^2} \, \frac{\phi_B^+(\omega)}{\omega}
- v_\perp^2 \, \frac{S_B^{ (i)}(u)}{\bar u} 
 \frac{\mu_\pi \phi_\sigma(v)}{6 E_2 \bar v^3} \, \frac{\phi_B^-(\omega)}{\omega}
\right. 
\cr & \qquad\phantom{ C_{FA}}
\left. 
+ v_\perp^2 \, \frac{S_B^{(i)}(u)+S_B^{(ii)}(u)}{\bar u} 
 \frac{ \mu_\pi \phi_\sigma(v)}{6 E_2 \bar v^3} \, \frac{\phi_B^+(\omega)}{\omega}
\right) \,.
\end{align}

\vspace{1em}

\subsubsection{Diagram (B6)}

\begin{align}
g_{(B6)}^{\rm finite} &= C_A \left( 
\frac{S_B^{(i)}(u)}{\bar u}
  \left( \left(1-\frac{2E_2}{M_B} \right) \frac{\phi_\pi(v)}{2v} 
    +v_\perp^2 \, \frac{\phi_\pi(v)}{2\bar v} \right) 
 -\frac{ E_2}{M_B} \, \frac{s_3}{\bar u} 
  \frac{\phi_\pi(v)}{2 v} \right) \frac{\phi_B^+(\omega)}{\omega}  
 \,,
\end{align}
and 
\begin{align}
g_{(B6)}^{\rm endpoint} &= 
C_A \left( \frac{2 E_2 M_B}{k^2} \, \frac{s_5}{\bar u^2} \,
\frac{(\bar v - v) \, \phi_\pi(v)}{ 4 v \bar v} \, \frac{\phi_B^+(\omega)}{\omega}
-  \frac{S_B^{\rm (i)}(u)+S_B^{\rm (ii)}(u)}{\bar u} 
\frac{\phi_\pi(v)}{2 v \bar v} \,  \frac{\phi_B^-(\omega)}{\omega} \right)
\,.
\end{align}
\clearpage 

\section{More on Kinematics}
\label{app:kinematics}

Expressing the energies $E_{1,2}$ of the two pions in terms of
the kinematic variables $k^2,q^2,\cos\theta$, one obtains
\begin{align}
  E_{1,2}(k^2,q^2,\cos\theta) &= \frac{k^2 + M_B^2 -q^2 \pm \cos\theta \, \sqrt{\lambda(k^2,q^2)}}{4 M_B} \,.
\end{align}
Without loss of generality, we may assume that $\cos\theta \geq 0$, such that
$E_2 < E_1$, and we thus have to determine
the minimal value of $E_2$ for given phase-space constraints on
$(k^2,q^2,\cos\theta)$,
\begin{align}
  E_{\rm min} &= \min E_2(k^2,q^2,\cos\theta) \qquad \mbox{(for $\cos\theta \geq 0$)} \,.
\end{align}
(For $\cos \theta \leq 0$, the same discussion goes through for $E_1$.)
Since $E_2$ is \emph{decreasing} with $\cos\theta$, its minimal value (for fixed $(k^2,q^2)$)
is obtained for the maximal value $\cos\theta|_{\rm max} \equiv 1/a $ with $a \geq 1$.
Similarly, $E_2$ is \emph{increasing} with $k^2$, such that its minimal value is
obtained for $k^2 = k^2_{\rm min}$.
Concerning the $q^2$-dependence (for fixed values $k^2=k^2_{\rm min}$ and $\cos\theta=1/a$),
the situation is more involved. The function $E_2(q^2)$ exhibits a minimum at
\begin{align}
 q_\star^2 &= M_B^2 + k^2_{\rm min} - \frac{2 a M_B \sqrt{k^2_{\rm min}}}{\sqrt{a^2-1}} \,.
\end{align}
This always fulfills $q_\star^2 \leq q^2_{\rm max} = (M_B - \sqrt{k^2_{\rm min}})^2$,
which is the upper phase-space boundary for $q^2$. However,
the condition $q_\star^2 \geq 0$ yields a non-trivial relation between $k^2_{\rm min}$
and $a$:
\begin{align}
  \mbox{minimum at $q_\star^2 \geq 0$} \quad \Leftrightarrow \quad k_{\rm min}^2 \leq \frac{a-1}{a+1} \, M_B^2 \,.
  \label{rel}
\end{align}
We thus have to consider two cases
\begin{itemize}
\item $q_\star^2 \geq 0$, with
\begin{align}
  E_{\rm min} & = E_2(k^2_{\rm min}, q_\star^2,1/a) = \frac{\sqrt{a^2-1}}{2a} \, \sqrt{k^2_{\rm min}}
  \cr
  \quad \Leftrightarrow \quad & k^2_{\rm min} = \frac{4a^2}{a^2-1} \, E_{\rm min}^2 \,,
  \label{rel1}
\end{align}
for which the relation (\ref{rel}) translates into (using $E_2 \leq M_B/2$)
\begin{align}
  E_{\rm min} < \frac{a-1}{a} \, \frac{M_B}{2} \,.
\end{align}

\item $q_\star^2 <0$, with
\begin{align}
E_{\rm min} & = E_2(k^2_{\rm min},0,1/a) = \frac{(a+1) \, k^2 + (a-1) \, M_B^2}{4 a M_B}
\cr
\quad \Leftrightarrow \quad & k^2_{\rm min} = \frac{4 a M_B \, E_{\rm min} - (a-1) \, M_B^2}{a+1}
\label{rel2}
\end{align}
for which the complement of the relation (\ref{rel}) now consistently translates into
\begin{align}
  E_{\rm min} > \frac{a-1}{a} \, \frac{M_B}{2} \,.
\end{align}
\end{itemize}

Notice that (\ref{rel2}) always holds for $a=1$, in which case the minimal value of $E_2$
is given at $q^2=0$, and $k^2_{\rm min} = 2 M_B \, E_{\rm min}$, as in Scenarios~A and B defined in the text.
For a given value of $E_{\rm min}$, there is a critical value of the angular cut,  
$a_* = M_B/(M_B - 2 E_{\rm min})$, above which (\ref{rel1}) is to be used.
In our Scenario~C we took $k^2_{\rm min} = M_B^2/4$ and $a=3$, for which one
actually has $q_\star^2 >0$, and therefore the correct expression for $k^2_{\rm min}$
reads
\begin{align}
& k^2_{\rm min} = M_B^2/4 \,, \quad |\cos \theta| \leq 1/3 \quad \cr & \Rightarrow \quad
  E_{\rm min} =  \frac{\sqrt{a^2-1}}{2a} \, \sqrt{k^2_{\rm min}} = \frac{1}{3\sqrt 2} \, M_B
  \simeq 1.24~{\rm GeV}\,.
\end{align}
(For the resulting value of $E_{\rm min}$ one has $a_*\simeq 1.89$,
and therefore $a > a_*$ in our Scenario~C.)

\bibliographystyle{JHEP}
\bibliography{references.bib}

\end{document}